\documentclass[12pt,a4paper]{JHEP3}
\preprint{Cavendish--HEP--10/04\\
MCnet/10/08}
\usepackage{cite}
\usepackage{amssymb}
\usepackage{amsmath}
\usepackage{epsfig}
\usepackage{subfigure}
\usepackage{color}
\let\normalcolor\relax
\usepackage{graphicx}
\usepackage{inputenc}
\usepackage{xspace}
\usepackage{slashed}
\inputencoding{latin1}
\usepackage{epstopdf}
\usepackage{axodraw4j}

\def\gev{~{\rm GeV}}

\def\bm{\boldmath}

\def\lsim{\mathrel{\raise.3ex\hbox{$<$\kern-.75em\lower1ex\hbox{$\sim$}}}}
\def\gsim{\mathrel{\raise.3ex\hbox{$>$\kern-.75em\lower1ex\hbox{$\sim$}}}}
\def\ifmath#1{\relax\ifmmode #1\else $#1$\fi}

\def\gtap{\lower .7ex\hbox{$\;\stackrel{\textstyle >}{\sim}\;$}}

\newcommand{\beq}{\begin{equation}}
\newcommand{\eeq}{\end{equation}}
\newcommand{\bea}{\begin{eqnarray}}
\newcommand{\eea}{\end{eqnarray}}
\newcommand{\met}{\not \!\! E_T}
\newcommand{\etam}{\eta_{\rm max}}

\newcommand{\alps}{\alpha_{\rm S}}

\title{Effects of invisible particle emission on global inclusive variables at hadron colliders\\}

\author{Andreas Papaefstathiou and Bryan Webber\\
Cavendish Laboratory, University of Cambridge,\\
J.J. Thomson Avenue, Cambridge, CB3 0HE, U.K.\\
Email: \email{andreas@hep.phy.cam.ac.uk, webber@hep.phy.cam.ac.uk}
}

\abstract{We examine the effects of invisible particle emission in
  conjunction with QCD initial state radiation (ISR) on quantities
  designed to probe the mass scale of new physics at hadron colliders,
  which involve longitudinal as well as transverse final-state
  momenta.  This is an extension of our previous
  treatment~\cite{Papaefstathiou:2009hp} of the effects of ISR
  on global inclusive variables.
  We present resummed results on the visible invariant mass
  distribution and compare them to parton-level Monte
  Carlo results for top quark and gluino pair-production at the
  LHC. 
 There is good agreement as long as the visible pseudorapidity
 interval is large enough ($\etam\gtap 3$).
 The effect of invisible particle emission is small in the case
  of top pair production but substantial for gluino pair production.
This is due mainly to the larger mass of the intermediate particles in
gluino decay (squarks rather than W-bosons).
We also show Monte Carlo modelling of the effects of
  hadronization and the underlying event.
The effect of the underlying event is large but may be approximately
universal.}


\begin{document}
\section {Introduction}
Amongst the many observables that could be of use in the search for
new physics at hadron colliders, those that do not depend on detailed
hypotheses about the structure of the final state may be the most suitable
for an initial general survey of the high-energy frontier.  Quantities
of this type have been investigated and named {\it global inclusive}
observables in ref.~\cite{Konar:2008ei}.   Such quantities can provide
information on the energy scales of new hard processes, for example
the production of exotic heavy particles, without imposing prejudices
about the nature of the new dynamics.

One may distinguish between global inclusive observables that depend
only on transverse momenta, such as the  visible and missing
transverse energies, and observables such as the visible energy and
invariant mass, which depend also on longitudinal momenta.  The former
are not  affected by the unknown motion of the hard process in the collider
frame, in the approximation that the process is initiated by partons
moving collinearly with the beams.  However, the colour charges of
those partons and the high scale of the hard process necessarily imply
that there is significant initial-state QCD radiation (ISR), which
contributes to both types of global inclusive observables.

In ref.~\cite{Konar:2008ei} various global inclusive observables were
investigated, including those that make use of longitudinal as well as
transverse momentum components.  The quantities studied included the
visible energy $E$, i.e.\ the sum of energies of particles registered
by the detector, and the visible invariant mass $M$,
\beq\label{eq:Mdef}
M = \sqrt{E^2-P_z^2-\met^2}\;,
\eeq
where $P_z$ is the visible longitudinal momentum.  Here $\met$,  the
missing transverse energy, is in practice defined as the negative of
the total visible transverse momentum, so that (\ref{eq:Mdef}) does
indeed define a Lorentz-invariant quantity.  In addition, in
\cite{Konar:2008ei} a new variable was introduced, defined as
\beq
\hat{s}^{1/2}_{\rm min}(M_{\rm inv}) \equiv \sqrt{M^2+\met^2}+\sqrt{M_{\rm inv}^2+\met^2}\ ,
\label{eq:smin_def}
\eeq
where the parameter $M_{\rm inv}$ is a variable estimating the sum of masses of all
invisible particles in the event:
\beq
M_{\rm inv} \equiv \sum_{i=1}^{n_{\rm inv}} m_i\ .
\label{eq:minv}
\eeq
It was argued that the peak in the distribution of $\hat{s}^{1/2}_{\rm
  min}$ is a good indicator of the mass scale of new physics processes
involving heavy particle production.

In ref.~\cite{Papaefstathiou:2009hp} we examined the effects of QCD
initial state radiation (ISR) on these observables, first in an
approximate fixed-order treatment, taking into account
collinear-enhanced terms, and then in an all-orders resummation of
such terms.  We quantified the way the distributions of quantities
that involve longitudinal momenta depend on the scale of the
underlying hard subprocess and on the properties of the detector, in
particular the maximum visible pseudorapidity $\etam$.

\begin{figure}
  \centering 
  \includegraphics[scale=0.70]{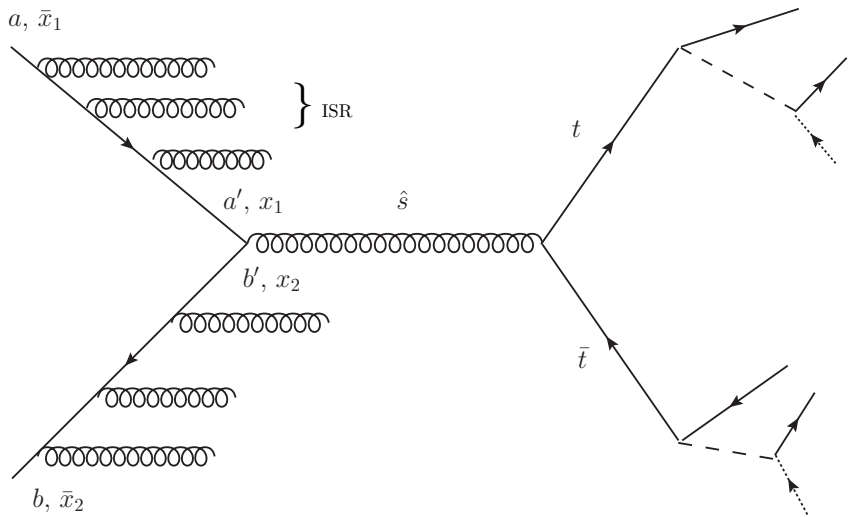}
  \caption{Feynman diagram for $qq'$ + ISR $\rightarrow g \rightarrow
  t\bar{t} \rightarrow jjll \slashed{E_T}$.The solid lines represent
  visible jets or leptons, the dotted lines invisible neutrinos.}
\label{fig:feynman}
\end{figure}

In the treatment in ref.~\cite{Papaefstathiou:2009hp} we supposed that
all the final-state particles from the hard subprocess are detected.
We considered in detail the case of top quark pair production when
both the $t$ and $\bar{t}$ decay hadronically. In the present paper we
study the effect of invisible particle emission, which allows us to
extend the validity of our treatment to many cases of Standard Model
and beyond Standard Model processes of interest at the LHC.  We
examine semi-leptonic/hadronic and fully semi-leptonic decays of the top quark (see
Fig.~\ref{fig:feynman}). We also consider gluino pair-production,
where the final decay products contain heavy invisible particles, the
lightest supersymmetric particles (LSPs), which remain undetected. The
emission of the LSPs in the decay of the gluino can
cause a sizeable effect on global inclusive observables. We
examine the interplay of this effect with ISR effects.

The Monte Carlo results presented in
refs.~\cite{Konar:2008ei,Papaefstathiou:2009hp} show that the second
term on the right-hand side of eq.~(\ref{eq:smin_def}) is not strongly
affected by ISR. The first term is intended to add extra longitudinal
information about the hard subprocess, allowing a more reliable
determination of its mass scale. The extra information enters through
the visible mass $M$. Therefore, as in
ref.~\cite{Papaefstathiou:2009hp}, we concentrate on this quantity. 

The paper is organised in the following way: In section~\ref{sec:isrorig}
we review the ISR resummation. Then in section~\ref{sec:isrinvis}
we develop our treatment of ISR effects to
include one or two invisible decays originating from the hard
process. In section~\ref{sec:invresults} we present comparisons of
resummed distributions to distributions produced using the
general-purpose Monte Carlo event generator \texttt{Herwig++}~\cite{Bahr:2008pv} for
$t\bar{t}$ and $\tilde{g}\tilde{g}$ production. Moreover, we examine
the effect of hadronization of ISR and the effect of the underlying
event on the parton-level distributions.  Our conclusions are summarized in
section~\ref{sec:conc}.  Appendix~\ref{app:mellin} contains details of
the numerical Mellin inversion method used to obtain the evolution
kernels and appendix~\ref{app:cross-sections} gives the parton-level
cross sections for the processes under investigation.

\section{ISR effects without invisible particle emission}\label{sec:isrorig}
We first review briefly the main results of
ref.~\cite{Papaefstathiou:2009hp}. Consider the emission of ISR
partons from incoming partons $a$ and $b$, as in
Fig.~\ref{fig:feynman}.  If all the products of the hard subprocess
and all the ISR emitted at angles greater than $\theta_c$ are detected,
the resummed differential cross section may be written as:
\beq\label{eq:Sigma}
M^2\frac{d\sigma_{ab}}{dM^2 dY} = \int dx_1\,dx_2\,K_{a'a}(x_1/\bar x_1)
f_a(\bar x_1,Q_c)\,K_{b'b}(x_2/\bar x_2)f_b(\bar x_2,Q_c)\,\hat\sigma_{a'b'}(x_1x_2S)\;\;,
\eeq
or equivalently
\beq\label{eq:SigmaDef}
S\frac{d\sigma_{ab}}{dM^2 dY} = \int dz_1\,dz_2\,K_{a'a}(z_1)
f_a(\bar x_1,Q_c)\,K_{b'b}(z_2)f_b(\bar x_2,Q_c)\,\hat\sigma_{a'b'}(z_1z_2M^2)\;\;,
\eeq
where $M$ and $Y$ are the invariant mass and rapidity of the detected
system (the `visible' mass and rapidity) and
\beq\label{eq:xbar}
\bar x_1 = \frac{M}{\sqrt S}e^{Y}\;,\;\;\;
\bar x_2 = \frac{M}{\sqrt S}e^{-Y}\;,
\eeq
$\sqrt S$ being the overall c.m.\ energy-squared.
The hard subprocess cross section $\hat\sigma_{a'b'}$ is evaluated at
c.m.\ energy squared $Q^2=x_1x_2S=z_1z_2M^2$, where $z_i=x_i/\bar
x_i$.  The parton distribution functions (PDFs) $f_{a,b}$ are evaluated
at the lower scale $Q_c\sim \theta_c Q$.\footnote{We find that results
  are somewhat sensitive to the constant of proportionality here.  We
  actually use $Q_c = Q\exp(-\etam)$ where $\etam=-\ln\tan(\theta_c/2)$
  is the maximum visible pseudorapidity.}
The kernel functions $K_{a'a}$ and $K_{b'b}$ describe the evolution of
the PDFs from scale $Q_c$ to $Q$.  They satisfy an
evolution equation like that of the parton distributions themselves:
\beq\label{eq:evolK}
Q\frac{\partial}{\partial Q}K_{b'b}(z) =\frac{\alps(Q)}\pi\int\frac{dz'}{z'} P_{b'a}(z') K_{ab}(z/z')\;,
\eeq
with the initial condition that $K_{ab}(z)=\delta_{ab}\delta(1-z)$ at $Q=Q_c$.  
We describe in appendix~\ref{app:mellin} the method that we used to compute the
kernel functions.

The main conclusions from the study in ref.~\cite{Papaefstathiou:2009hp}
are that, in the absence of invisible particles, the above analytical
results are in good agreement with those of Monte Carlo simulations, and
that the distribution of the new variable (\ref{eq:smin_def}) is
indeed determined primarily by that of the visible mass $M$.

\section{ISR effects including invisible particle emission}\label{sec:isrinvis}
Suppose now that an invisible 4-momentum $p^{\mu}_{inv}$ is emitted from
the hard subprocess. If we define the total lab-frame 4-momentum of the incoming
partons $a$ and $b$ as $P^{\mu} = (E,\vec{P})$, 
\begin{equation}\label{eq:total4mom}
P^{\mu} = \frac{1}{2} \sqrt{S} [ (\bar{x}_1 + \bar{x}_2), 0, 0, (\bar{x}_1 - \bar{x}_2) ] \;,
\end{equation}
 then the visible 4-momentum will be $P^{\mu} - p^{\mu}_{inv}$.
By definition, the visible mass is then given by:
\begin{equation}\label{eq:vismassdef}
M^2 = (P -  p_{inv})^2 = P^{\mu} P_{\mu} + p^{\mu}_{inv} p_{inv,\mu} - 2 p^{\mu}_{inv} P_{\mu}\;.
\end{equation}
Equation~(\ref{eq:vismassdef}) demonstrates the interplay between two effects: on one hand ISR increases the `true' scale of the hard process $Q$, to the `apparent' scale $M$ by contaminating the detector with extra particles, and on the other hand the invisible particle emission decreases $M$ by the loss of particles. In the case of gluino pair-production both effects are equally important, as we will show. 

Substituting from eq.~(\ref{eq:total4mom}) in eq.~(\ref{eq:vismassdef}) and defining
$p^{\pm}_{inv} = p^0_{inv} \pm p^3_{inv}$  we obtain
\begin{equation}\label{eq:vismass3}
M^2 = \bar{x}_1 \bar{x}_2 S + m_{inv}^2 - \sqrt{S} [ \bar{x}_1 p^-_{inv} + \bar{x}_2 p^+_{inv} ]\;,
\end{equation}
where $m_{inv}$ represents the total invariant mass of the invisibles,
$m_{inv}^2 = p^{\mu}_{inv} p_{inv,\mu}$.

The momenta $p^\mu_{inv}$ are defined in the lab frame, relative to which
the c.m.\ frame of the hard subprocess is boosted by an
amount defined by the momentum fractions $x_1$ and $x_2$ of the
partons entering the subprocess. This implies that the $p^{\pm}_{inv}$ transform as:
\begin{eqnarray}\label{eq:pplusminus}
p^+_{inv} = \sqrt{\frac{x_1}{x_2}} q^+_{inv} \;,\;\;\; p^-_{inv} = \sqrt{\frac{x_2}{x_1}} q^-_{inv}\;,
\end{eqnarray}
where $q^{\pm}_{inv} = q^0_{inv} \pm q^3_{inv}$, defined in terms of the invisible momentum, $q^{\mu}_{inv}$, in the c.m.\ frame of the hard subprocess. Substituting the expressions of eq.~(\ref{eq:pplusminus}) in eq.~(\ref{eq:vismass3}) we find an expression for the visible invariant mass:
\begin{eqnarray}
M^2 = m_{inv}^2 + \bar{x}_1 \bar{x}_2S \left[  1 - z_1 f^+_{inv} - z_2 f^-_{inv} \right]\;,
\label{eq:invmsq}
\end{eqnarray}
where we have defined $f^{\pm}_{inv} = q^{\pm}_{inv}/Q$ and
used $Q^2 = \bar{x}_1 \bar{x}_2 z_1 z_2 S$. We may now solve
eq.~(\ref{eq:invmsq}) for $Q^2$ to obtain $Q^2$ in terms of $M^2$:
\begin{eqnarray}\label{eq:s}
Q^2 = \frac{ z_1 z_2 (M^2 - m_{inv}^2)}{1 - z_1 f^+_{inv} - z_2 f^-_{inv}}\;.
\end{eqnarray}
The above expression for the hard subprocess scale now becomes the argument
of the parton-level cross section, $\hat{\sigma}_{a'b'}$ in eq.~(\ref{eq:SigmaDef}):
\beq\label{eq:SigmaDefinv}
S\frac{d\sigma_{ab}}{dM^2 dY} = \int dz_1\,dz_2\,K_{a'a}(z_1)f_a(\bar x_1,Q_c)\,K_{b'b}(z_2)f_b(\bar x_2,Q_c) 
\,\hat\sigma_{a'b'}\left( \frac{ z_1 z_2 (M^2 - m_{inv}^2)}{1 - z_1 f^+_{inv} - z_2 f^-_{inv}} \right)\,.
\eeq
The functions $f^{\pm}_{inv}$, which are related to the invisible
particle four-momenta, remain to be determined. The visible system
rapidity, $Y$, is also modified by the presence of invisible particles as:
\begin{equation}\label{eq:Yinv}
Y = \frac{1}{2} \log \left( \frac{\bar{x}_1 ( 1 - z_1 f^+_{inv} )} { \bar{x}_2 ( 1 - z_2 f^-_{inv} ) } \right)\;,
\end{equation}
and therefore eqs. (\ref{eq:xbar}) for $\bar{x}_{1,2}$ become
\bea\label{eq:xbarinv}
\bar x_1 &=& \sqrt{\frac{(M^2 - m_{inv}^2)(1 - z_2 f^-_{inv})}{S(1 -
   z_1 f^+_{inv} - z_2 f^-_{inv})(1- z_1 f^+_{inv})}} e^{Y}\;,\nonumber\\
\bar x_2 &=& \sqrt{\frac{(M^2 - m_{inv}^2)(1 - z_1 f^+_{inv})}{S(1 -
   z_1 f^+_{inv} - z_2 f^-_{inv})(1- z_2 f^-_{inv})}} e^{-Y}\;.
\eea
The kinematic constraints restrict $Q^2$ to be greater than the threshold energy squared for the process and the true invariant mass, $M_{true}^2 \equiv \bar{x}_1 \bar{x}_2 S = Q^2 / (z_1 z_2)$, to be greater than the visible invariant mass, $M^2$. These result in the following constraints for $Q^2$:
\begin{equation}
Q^2 > Q_{threshold}^2 \;,\;\;\; Q^2 > z_1 z_2 M^2 \;.
\end{equation}

\subsection{Single-invisible decays}\label{sec:1inv}
The benchmark scenario for a single invisible decay originating from the hard process is $t\bar{t}$ production in which one of the two tops decays into $bqq'$ (hadronic) and the other into $b\ell\nu$ (semi-leptonic), the neutrino comprising the missing four-momentum. Excluding the proton remnants, we assume that all other particles within the pseudorapidity coverage are detected. We will refer to the neutrino as the invisible particle and the $W$ as the intermediate particle in the $t\bar{t}$ case, but the treatment is readily applicable to the gluino case where the invisible particle is the $\chi_1^0$ and the intermediate particle is a squark (treated in section~\ref{sec:2inv}). 

\begin{figure}\label{fig:decchain}
 \begin{centering}
   \includegraphics[scale=0.70]{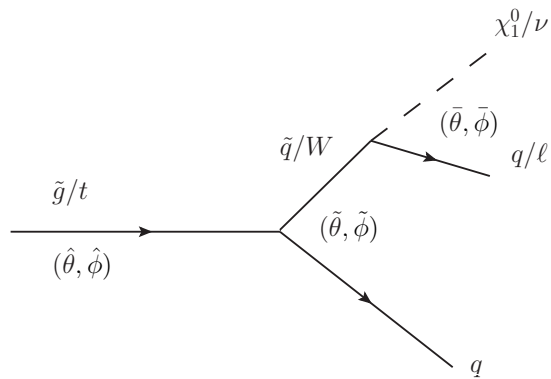}
   \caption{The sequential two-body decay chain under consideration in
     the invisible particle treatment. The relevant production angles
     in the parent centre-of-mass frame are also shown in parentheses.}
\end{centering}
\end{figure}

To calculate the functions $f^{\pm}_{inv}$ and obtain $Q^2$, we need to calculate the neutrino four-momentum in the hard process frame. This is done by choosing the neutrino four-momentum in the frame of its parent $W$ and then applying two subsequent Lorentz boosts: one going from the $W$ frame to the top frame, and one from the top frame to the hard process frame. The decay chain is shown in Fig.~\ref{fig:decchain}. Each of these boosts involves two angular variables which originate from the `decay' of the parent particle. Hence the four momentum $q^{\mu}_{inv}$ of the neutrino may be written as
\begin{equation}\label{eq:qinv}
q^{\mu}_{inv} = \Lambda^{\mu}_\kappa\left(Q, \hat{\theta}, \hat{\phi} \right) \Lambda^{\kappa}_\lambda \left( \tilde{\theta} , \tilde{\phi} \right) \bar{p}^\lambda_\nu(\bar{\theta}, \bar{\phi})\;,
\end{equation}
where the $\Lambda$'s are Lorentz boost matrices and where quantities
with a hat refer to the hard process frame, quantities with a tilde
refer to the top frame and quantities with a bar refer to the $W$
frame. The angles $\theta$ and $\phi$ represent the usual polar
angles, defined with respect to the direction of the `sister'
particle (see Fig.~\ref{fig:decchain}).
For example, in the case $W^+ \rightarrow \ell ^+ \nu_\ell$,
where the $W^+$ was produced from the top decay along with a bottom
quark, the angles $(\bar{\theta},\bar{\phi})$ are defined with respect
to the direction of motion of the $b$ in the $W^+$ frame. The two
boost vectors have magnitudes given by $| \vec{\beta}_i| = |\vec{p}_i|
/ E_i$ ($i=t,W$), the ratio of the parent 3-momentum magnitude and its energy. The boosts, as well as the magnitude of the invisible particle four-momentum, can obtained by considering kinematics in each frame as:
\begin{align}
&\bar{p}^\lambda_\nu(\bar{\theta}, \bar{\phi}) = \frac{m_W}{2}  ( 1, \vec{\bar{r}}) \;,\\\nonumber
&\vec{\beta}_W = \frac{m_t^2-m_W^2}{m_t^2 + m_W^2}  \vec{\tilde{r}} \;,\\\nonumber 
&\vec{\beta}_t = \sqrt{ 1 - \frac{4 m_t^2}{Q^2}}  \vec{\hat{r}} \;,
\end{align}
where $\vec{r} = (\sin\theta \cos \phi, \sin\theta \sin \phi, \cos
\phi)$ is the unit vector in spherical polar coordinates in the
appropriate frame and $m_{W}$, $m_t$ are the $W$ and top quark masses respectively. The four-vector $f^\mu_{inv}$, and hence the functions $f^{\pm}_{inv}$, are calculated by $f^{\pm}_{inv} = q^{\pm}_{inv}/Q$. Evidently, the functions $f^{\pm}_{inv}$ are functions of $Q^2$, giving an implicit equation for $Q^2$. To make this more explicit, we re-write eq.~(\ref{eq:s}):
\begin{equation}\label{eq:implicits}
Q^2 = \frac{ z_1 z_2 \left[M^2 - m_{inv}^2(Q^2, \Omega)\right]}{1 - z_1 f^+_{inv}(Q^2, \Omega) - z_2 f^-_{inv} (Q^2, \Omega)} \;,
\end{equation}
and analogously for eq.~(\ref{eq:xbarinv}),
where $\Omega$ represents the set of all angular
variables. In the present case $m_{inv}(Q^2, \Omega)=m_\nu\simeq 0$
but for multiple invisible particles it will also be a function as
indicated.

Equation (\ref{eq:implicits}) needs to be solved numerically for
each set ($z_1, z_2, \Omega$) in the region $(4m_{t / \tilde{g}}^2,
z_1 z_2 S)$, where $S$ is the square of the proton centre-of-mass
energy, along with the restriction that the visible invariant mass
should be lower than the `true' invariant mass, $M \leq M_{true}$. The
numerical solution was found using the Van Wijngaarden-Dekker-Brent
method~\cite{brent, gslmanual}, a bracketing method for finding roots
of one-dimensional equations. Since $Q$ is not uniquely determined for
each $M$, different values of the `true' centre-of-mass energy $Q$
contribute to the cross section. Note that not all possible
configurations ($z_1, z_2, \Omega$) are kinematically allowed to
contribute to the cross section at $M$ and hence some configurations
do not yield roots of eq.~(\ref{eq:implicits}). 
Once $Q^2$ is obtained, the parton-level cross section for the hard process partons, $\hat{\sigma}_{a'b'}(Q^2)$, is calculated. This result is then multiplied with the parton density functions for the incoming partons, $f_{a,b}(\bar{x}_{1,2}, Q_c)$, and the kernels for evolution from incoming partons $a$ and $b$ to hard process partons $a'$ and $b'$ ($K_{a'a}(z_1)$ and $K_{b'b}(z_2)$). We then integrate over all possible values of $z_1$ and $z_2$, according to eq.~(\ref{eq:SigmaDefinv}). Finally, to obtain the full resummed result we have to integrate over the distribution of the angular variables $\Omega$.
Notice that the visible invariant mass distribution becomes non-zero below the
threshold for production, $M < 2 m_ {t / \tilde{g}}$, owing to the
loss of invisible particles.

\subsection{Double-invisible decays}\label{sec:2inv}
We now turn to the case where both particles produced in the hard
process decay invisibly. For illustration we refer to sequential decays of the gluino: $\tilde{g} \rightarrow \tilde{q} q \rightarrow \chi^0_1 qq$. Although this decay mode is generally not the dominant one, it is useful for illustration of the procedure. We extend the treatment given in the semi-leptonic/hadronic top case by writing out functions related to the two invisible particle four-momenta in the decay chain (which we call $\chi$ and $\chi '$),
\begin{eqnarray}\label{eq:qnuqnubar}
q^{\mu}_{\chi} = \Lambda^{\mu}_\kappa\left(Q, \hat{\theta}, \hat{\phi} \right) \Lambda^{\kappa}_\lambda \left( \tilde{\theta} , \tilde{\phi} \right) \bar{p}^\lambda_\chi(\bar{\theta}, \bar{\phi})\;,\\
q^{\mu}_{\bar{\chi '}} = \Lambda^{\mu}_\kappa\left(Q, \hat{\theta'}, \hat{\phi'} \right) \Lambda^{\kappa}_\lambda \left( \tilde{\theta'} , \tilde{\phi'} \right) \bar{p}^\lambda_{\chi '}(\bar{\theta'}, \bar{\phi'})\;,
\end{eqnarray}
where the primed quantities now distinguish between the two
invisibles. Since both of these four-vectors are defined in the
hard subprocess frame, we have simply
\begin{equation}\label{eq:finv2}
f^{\pm}_{inv} = \frac{1}{Q}\left( q^{\pm}_{\chi} + q^{\pm}_{\chi '} \right)\;.
\end{equation}
The rest of the treatment is identical to the one-invisible case: an
implicit equation has to be solved to obtain $Q^2$ for each ($z_1$,
$z_2$, $\Omega$) set and then an 
integral over $\Omega$ is taken to obtain the resummed result.

\subsection{Angular distributions}\label{sec:angular}
\begin{figure}[htb]
  \centering 
  \vspace{1.0cm}
  \hspace{2.9cm}
  \begin{picture}(300,120)
  \put(0,0){\includegraphics[scale=0.34, angle=90]{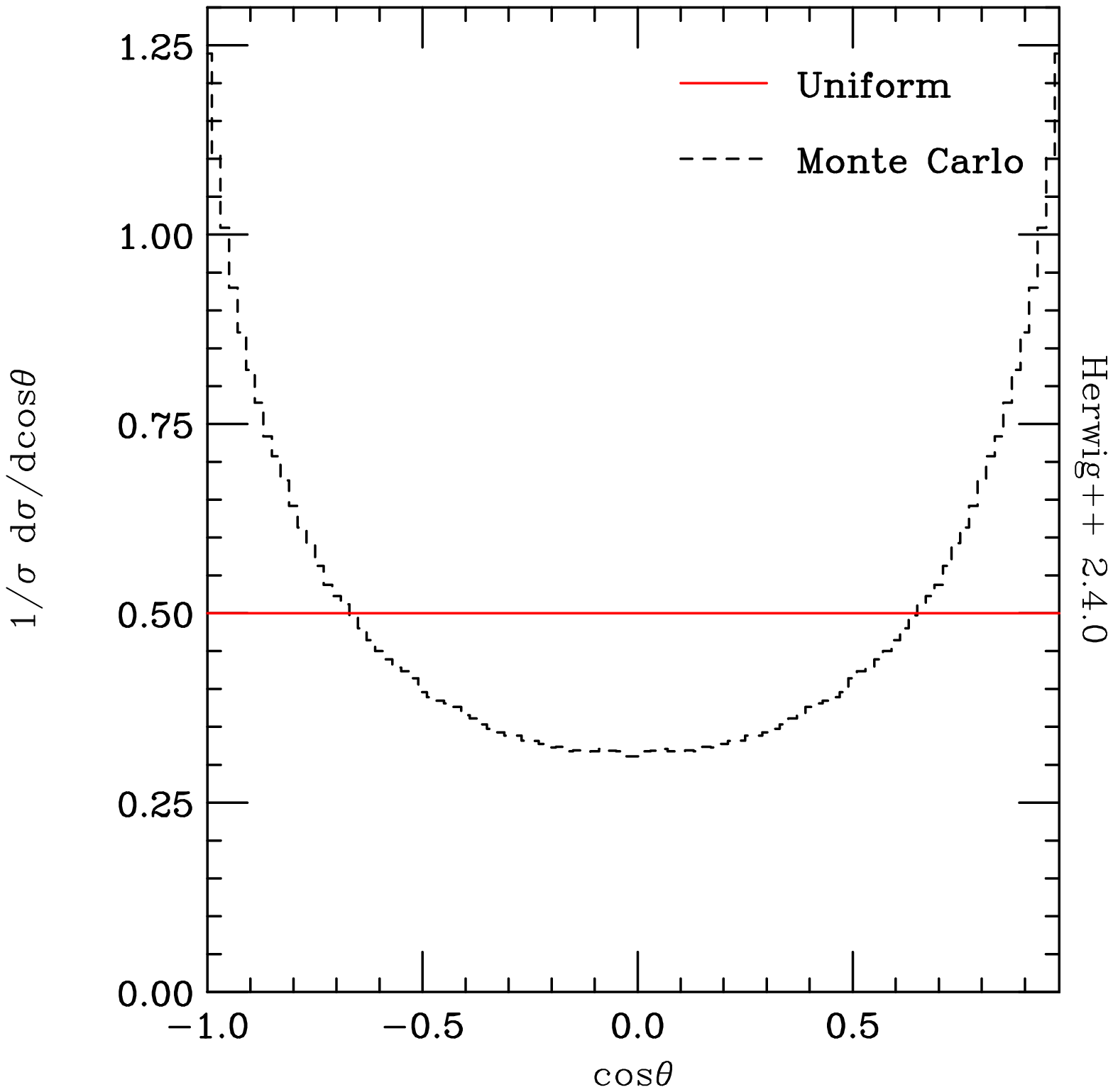}}
  \put(150,0){\includegraphics[scale=0.34, angle=90]{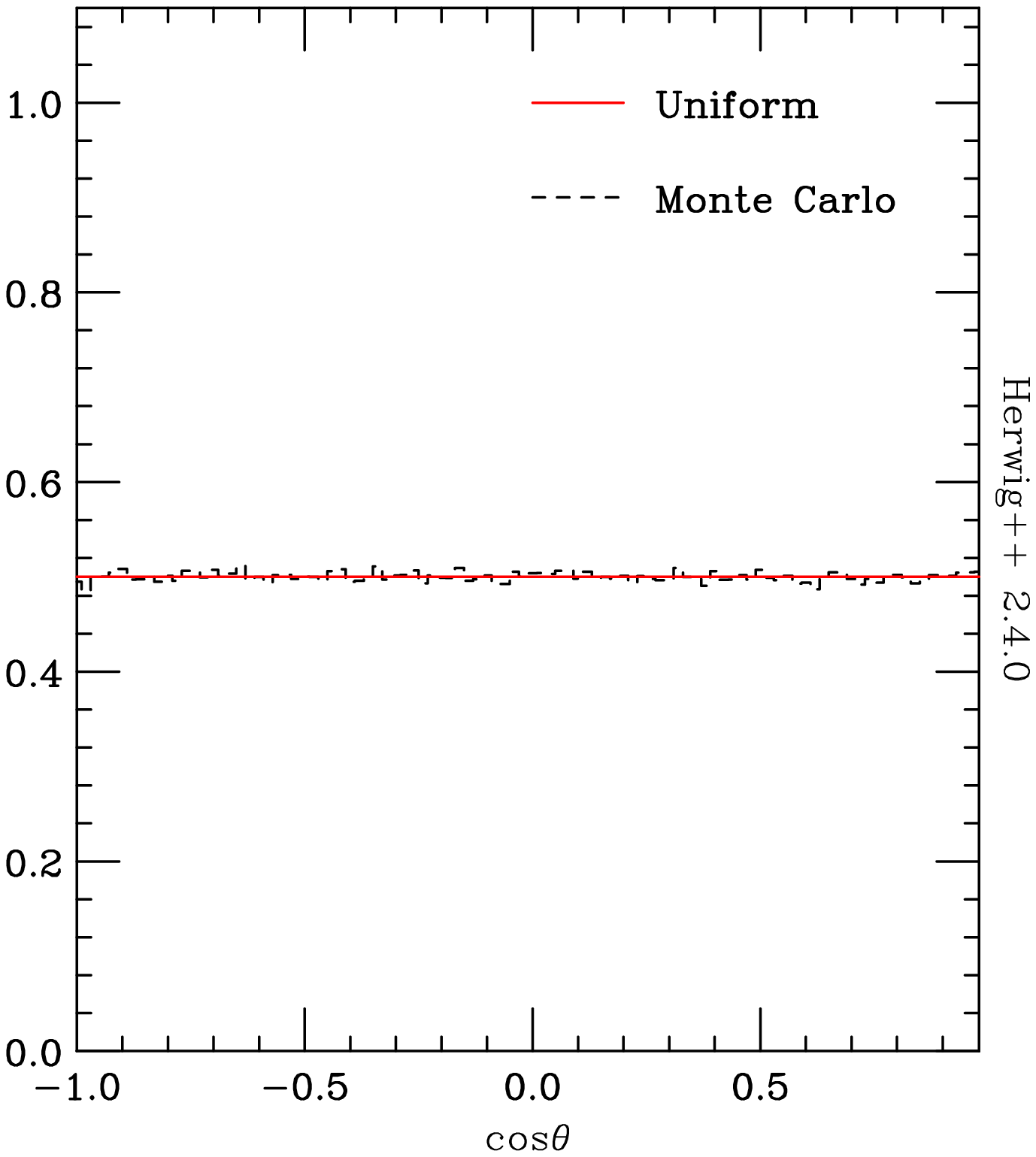}}
  \put(300,0){\includegraphics[scale=0.34, angle=90]{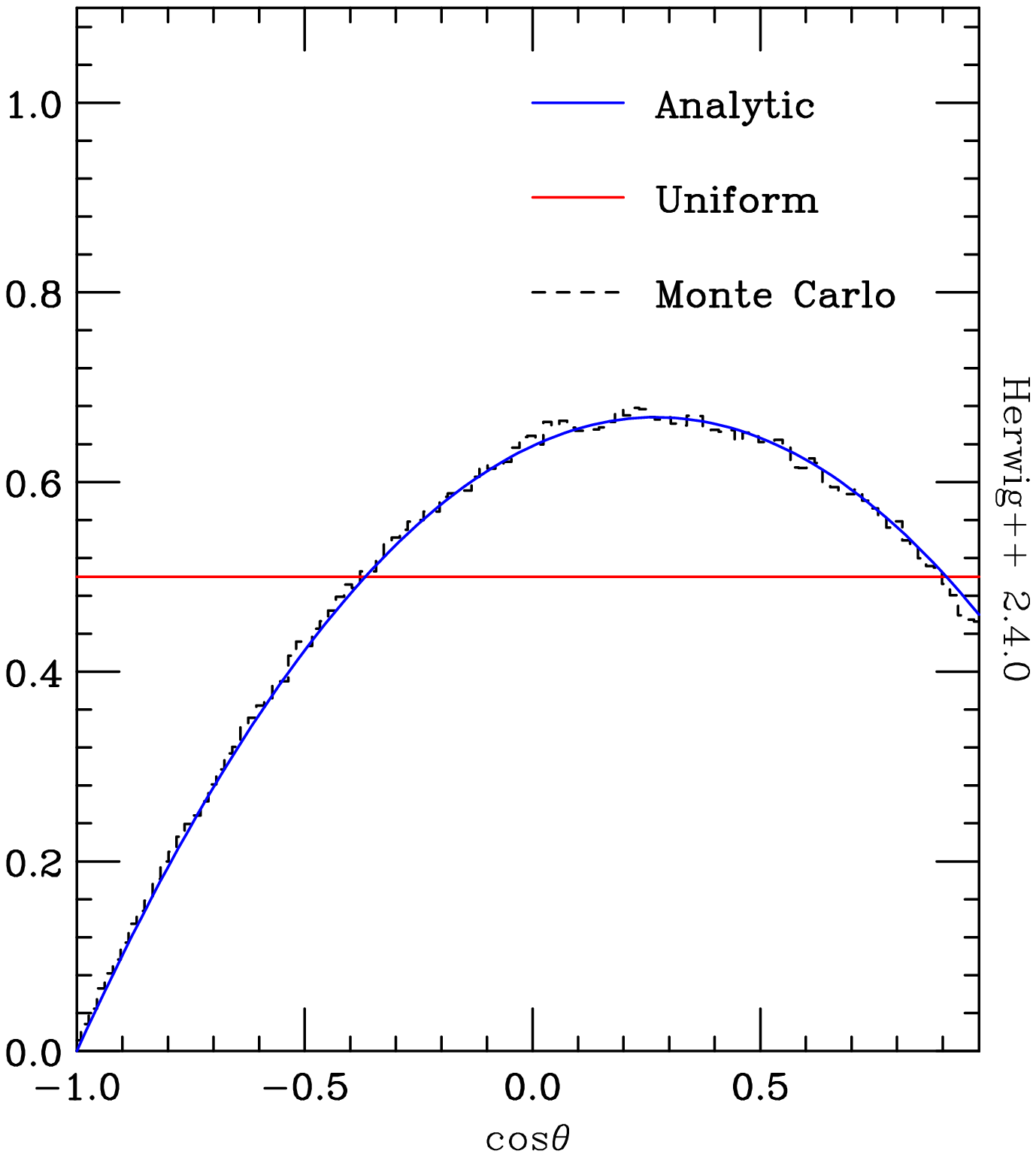}}
  \put(-25.5,-8.5){\small{$\hat{}$}}
  \put(124.5,-8.5){\small{$\tilde{}$}}
  \put(274.5,-8.5){\small{$\bar{}$}}
  \end{picture}
\caption{Monte Carlo results for the top pair-production decay chain angles. From left to right: the production angle of the top in the hard process frame, the angle of the outgoing $W$ boson in the top frame and the angle of the outgoing neutrino in the $W$ frame. The uniform distributions are shown for comparison. The neutrino angle in the $W$ frame is also compared to the analytic calculation. }
\label{fig:ttangles}
\end{figure}

The distributions of the angular variables $\Omega = (\hat{\theta},\hat{\phi}, \tilde{\theta}, \tilde{\phi}, \bar{\theta}, \bar{\phi})$, appearing in the treatment of invisibles given in the previous sections, are process-dependent. They represent the angles at which the daughter particle is emitted in the frame of the parent particle. We investigated the angular distributions using a Monte Carlo event generator (\texttt{Herwig++} 2.4.0~\cite{Bahr:2008pv}) and subsequently used the results in calculating the $f^{\pm}_{inv}$ functions. The results for SPS1a gluino pair-production are shown in Fig.~\ref{fig:ggangles}, where the uniform distributions are shown for comparison (red horizontal line). Figure~\ref{fig:ttangles} shows the distributions as obtained for top pair-production. The neutrino angle in the $W$ frame is also compared to the analytic calculation. As expected, all the $\phi$ angles, in both cases, were found to be uniform (not shown). The form of all the distributions can be justified using general spin considerations:

\begin{itemize}
\item[$\hat{\theta}_i$:] The angular distribution of the angle $\hat{\theta} _i$ at which the fermions are produced in the hard process frame is expected to have the form $\sim 1 + \beta \cos ^2 \hat{\theta}_i$, where $\beta$ is a process-dependent constant.

\item[$\tilde{\theta}_i$:] The angle $\tilde{\theta} _i$, is defined between the direction of the daughter boson ($W$ or $\tilde{q}$) with respect to the direction of polarization of the parent ($t$ or $\tilde{g}$) polarization. The angular distribution for a spin-up fermion parent is then given by~\cite{Hubaut:2005er}:
\begin{equation}
\frac{1}{N_\uparrow} \frac{\mathrm{d}N_\uparrow}{\mathrm{d} \cos \tilde{\theta} _i} = \frac{1}{2} ( 1 + P \alpha _i \cos \tilde{\theta} _i ) \;\;,
\end{equation}
where $\alpha _i$ is a constant and $P$ is the modulus of the polarization of the parent. Since the production processes for both $t\bar{t}$ and $\tilde{g} \tilde{g}$ are parity conserving, there is also an equal spin-down ($N_\downarrow$) contribution to the total distribution with the sign of $\alpha _i$ reversed. This results in a uniform distribution for $\cos \tilde{\theta}_i$. 
\begin{figure}[htb]
  \centering 
  \vspace{1.0cm}
  \hspace{2.9cm}
  \begin{picture}(300,120)
  \put(0,0){\includegraphics[scale=0.34, angle=90]{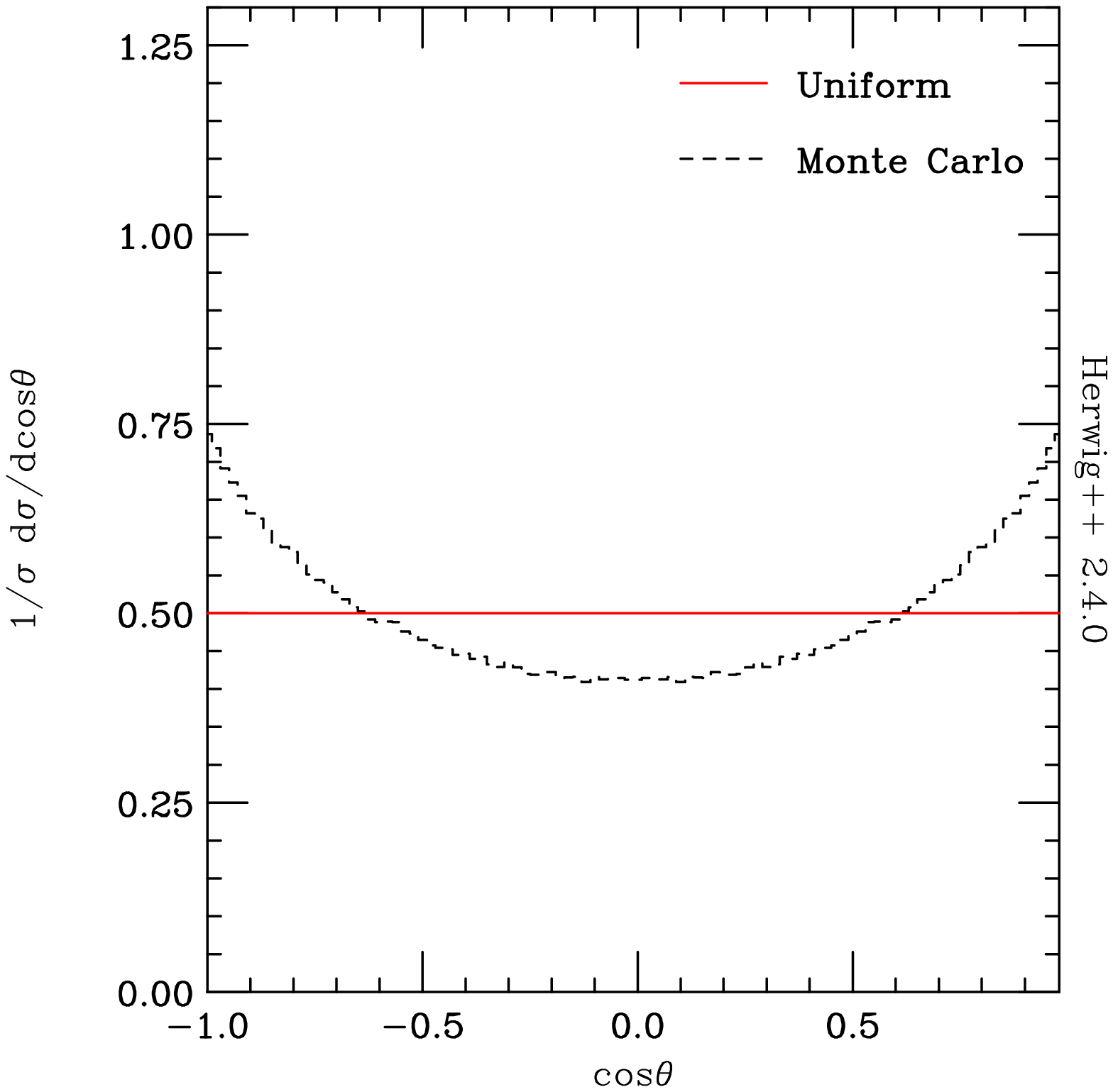}}
  \put(150,0){\includegraphics[scale=0.34, angle=90]{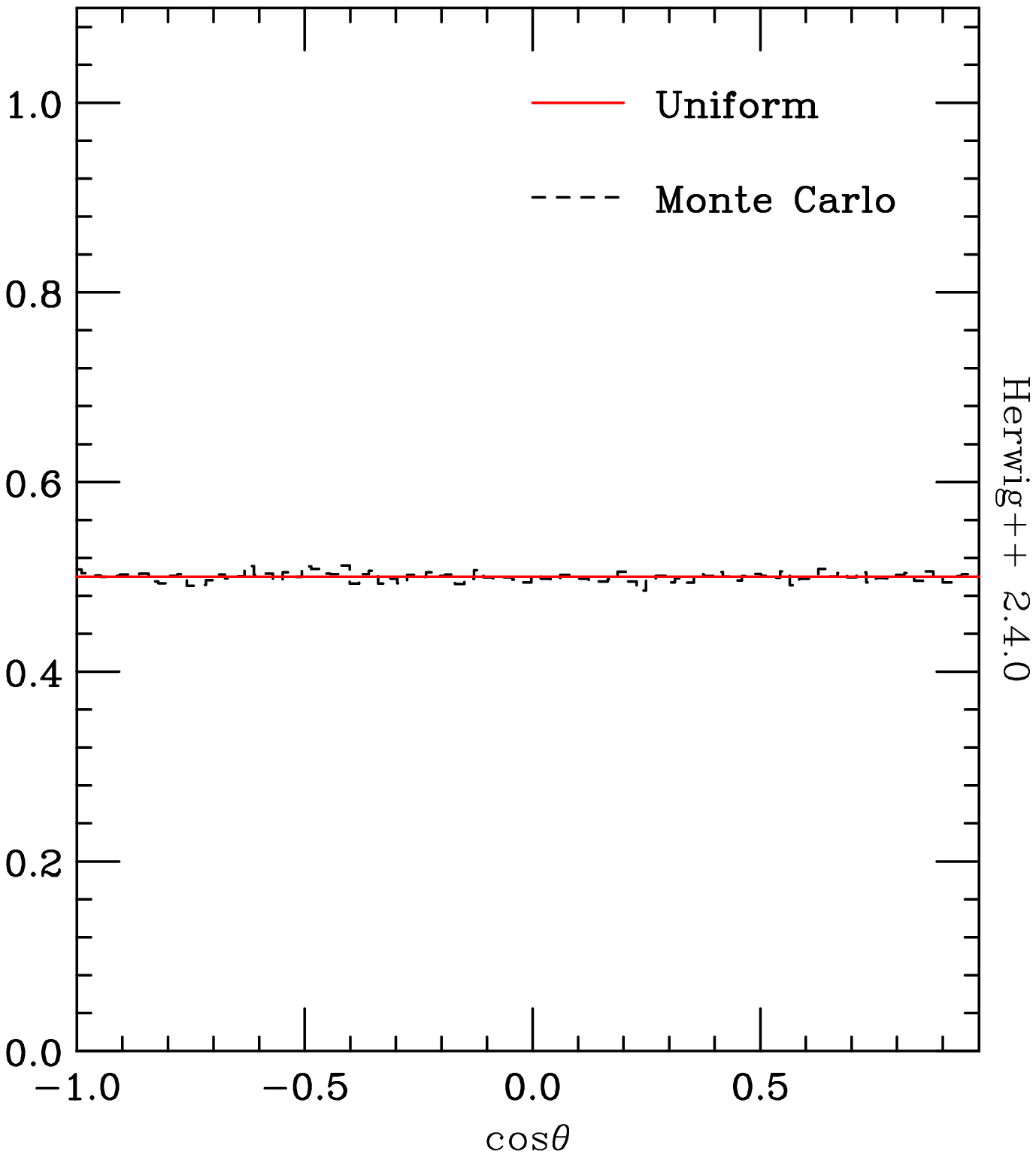}}
  \put(300,0){\includegraphics[scale=0.34, angle=90]{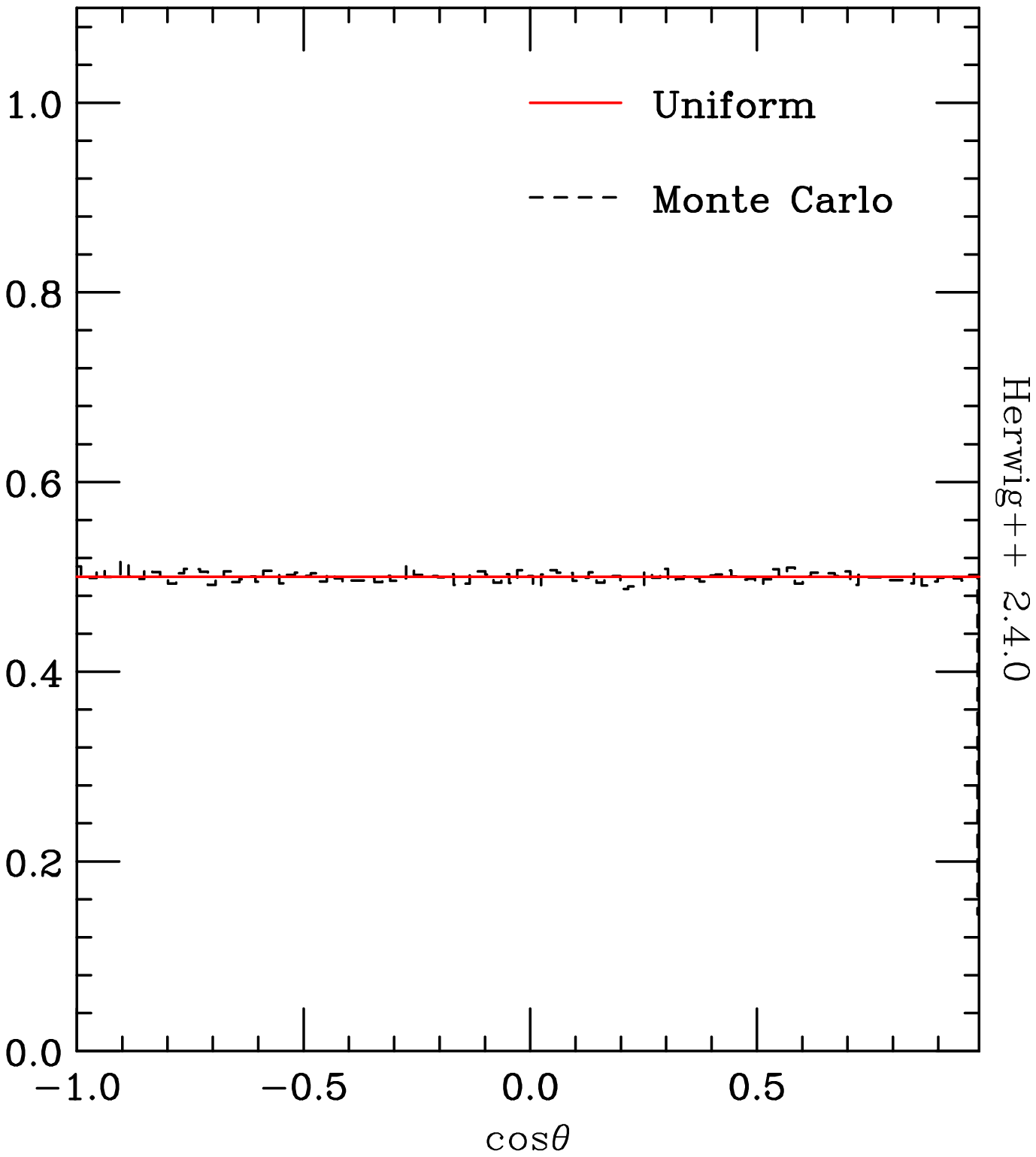}}
  \put(-25.5,-8.5){\small{$\hat{}$}}
  \put(124.5,-8.5){\small{$\tilde{}$}}
  \put(274.5,-8.5){\small{$\bar{}$}}
  \end{picture}
\caption{Monte Carlo results for the gluino pair-production decay chain angles. From left to right: the production angle of the gluino in the hard process frame, the angle of the outgoing squark in the gluino frame and the angle of the outgoing neutralino in the squark frame. The uniform distributions are shown for comparison.}
\label{fig:ggangles}
\end{figure}
\item[$\bar{\theta}_i$:] In gluino pair-production, the decay products
  of the squark, $\tilde{q}$, which is a scalar, are uniformly distributed in $\cos
  \bar{\theta}$.  In top pair-production, on the other hand, the decay
  $W\rightarrow \ell \nu _{\ell}$ is parity-violating and the
  distribution of $\cos \bar{\theta}$ is forward-backward asymmetric
  in the $W$ frame~\cite{qcdcollider}.  The angle $\bar{\theta}$
  (sometimes called $\Psi$, see e.g.~\cite{Aad:2009wy}) is used
  experimentally to infer helicity information on the $W$. The
  distribution may be written as
\begin{eqnarray}
\frac{1}{N} \frac{\mathrm{d} N} { \mathrm{d} \cos \bar{\theta} } = \frac{3}{2} \left[ F_0 \left( \frac{ \sin \bar{\theta} } { \sqrt{2} } \right)^2 + F_L \left( \frac{ 1 - \cos \bar{ \theta} } { 2} \right) ^2 +  F_R \left( \frac{ 1 + \cos \bar{ \theta} } { 2} \right) ^2  \right]\;\;,\nonumber \\
\end{eqnarray}
where $F_L$, $F_R$ and $F_0$ are the probabilities for left-handed,
right-handed and longitudinal helicities of the $W$ in top quark decay
respectively. The SM predictions, $(F_L, F_R, F_0) = (0.304, 0.001,
0.695)$, yield the curve shown on the right in Fig.~\ref{fig:ttangles}.
\end{itemize}

The spins of the two produced fermions (tops or gluinos) are correlated and
this may cause a degree of correlation between the distributions of
particles in the decay chains. We investigated whether these
correlations play an important role in the calculation of the
invisible particle effects on the visible mass. By comparing the
invariant mass distributions with and without the spin correlations in
the Monte Carlo we concluded that the effect is small in both top and
gluino pair-production and can be safely neglected.

\section{Results}\label{sec:invresults}
We present the resummed distributions obtained for $t\bar{t}$ and
$\tilde{g}\tilde{g}$ production according to
eq.~(\ref{eq:SigmaDefinv}). All results are for the LHC at design
energy, i.e.\ $pp$ collisions at $\sqrt s = 14$ TeV.  We have
integrated over the visible system rapidity, $Y$, in the range
$|Y|<5$. We first compare our results to those obtained using the
\texttt{Herwig++} event generator at parton level (i.e. no
hadronization or underlying event) and excluding the proton
remnants.\footnote{We  verified using the event generator that the
  contribution of the proton remnants to the total invariant mass in
  the considered rapidity range is negligible.}  In
sections~\ref{sec:hadronization} and \ref{sec:MPI} we examine the
effects of hadronization and the underlying event. Parton-level top
and gluino pair-production cross section formulae are given in
appendix~\ref{app:cross-sections}. The PDF set used both in the
calculation and \texttt{Herwig++} is the MRST LO** (MRSTMCal)
set~\cite{Sherstnev:2007nd, Sherstnev:2008dm}.

\subsection{Top quark pair production}\label{sec:topresults}
\begin{figure}
  \centering 
  \vspace{0.5cm}
    \includegraphics[scale=0.34, angle=90]{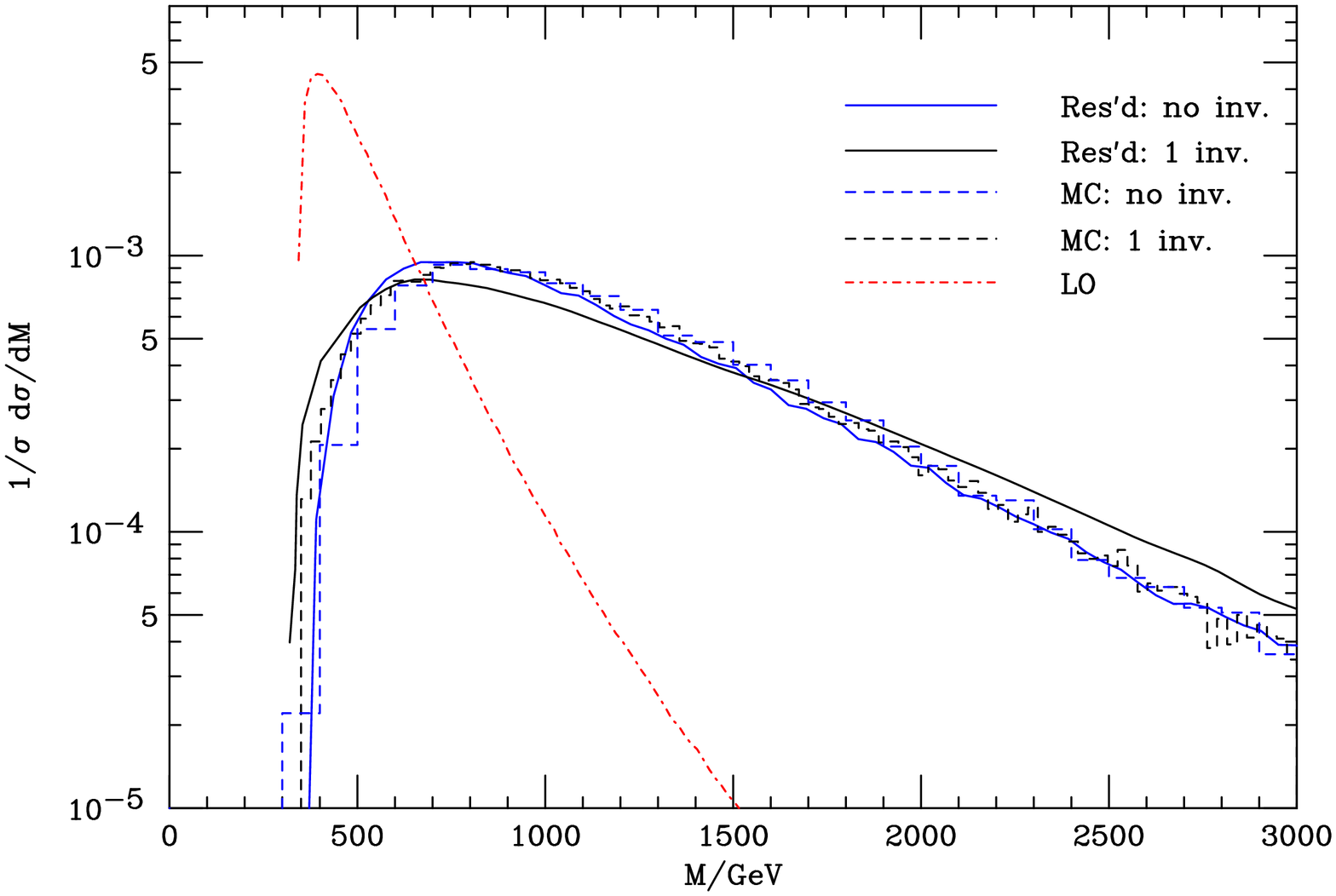}
    \hspace{0.5cm}
    \includegraphics[scale=0.34, angle=90]{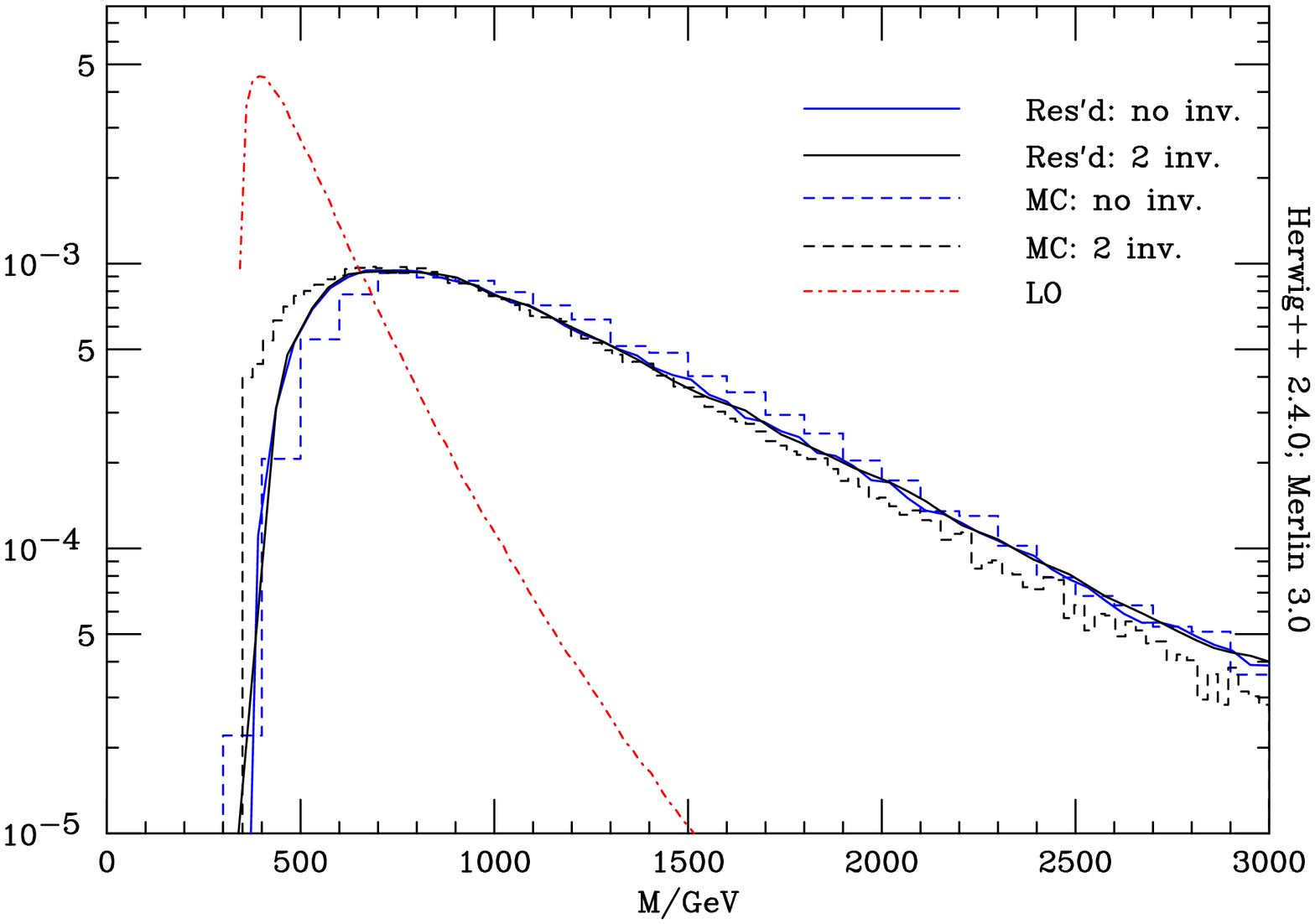}
\caption{The $t\bar{t}$ visible mass distributions for a pseudorapidity cut $\etam=5$. Left: comparing
hadronic (no invisibles) and semi-leptonic (one invisible) decays. Right: comparing
hadronic (no invisibles) and fully leptonic (two invisibles) decays.
The leading order $t\bar{t}$ invariant mass distribution is shown (red dot dashes) for comparison.}
\label{fig:tte5}
\end{figure}

We present resummed results in comparison to Monte Carlo for Standard
Model $t\bar{t}$ production, where we include particles with maximum
pseudorapidity $\etam = 5$. In Fig.~\ref{fig:tte5} we show
separate results for combinations of hadronic and semi-leptonic decays
of the top, leading to zero, one or two invisible neutrinos from the
hard process. The effect of the invisibles in both the fully
semi-leptonic case and the hadronic/semi-leptonic case are small
compared to the effects of hadronization, to be discussed in
section~\ref{sec:hadronization}. The differences between the Monte
Carlo and resummed curves in Fig.~\ref{fig:tte5} may be attributed to sensitivity to the
behaviour of the PDFs and parton showering at low
scales, and the precise definition of $Q_c$ in terms of $\etam$, since
$Q_c$ can be as low as $ 2 m_t \times e^{-5} \sim 2\gev$ in the case
of $t\bar{t}$ production.

\subsection{Gluino pair production}\label{sec:gluinoresults}
\begin{figure}
  \centering 
  \vspace{1.2cm}
  
    \includegraphics[scale=0.34, angle=90]{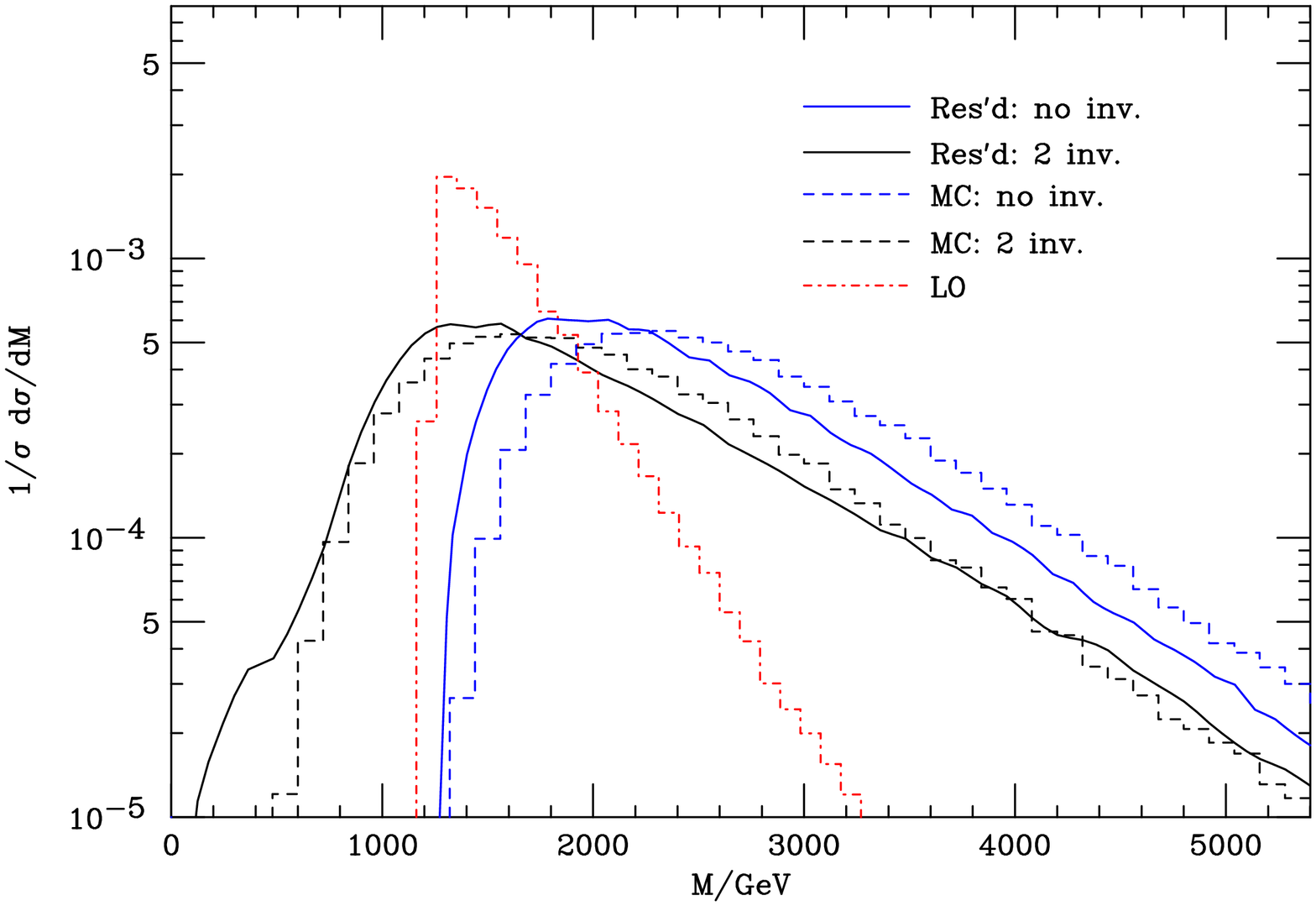}
    \hspace{0.5cm}
    \includegraphics[scale=0.34, angle=90]{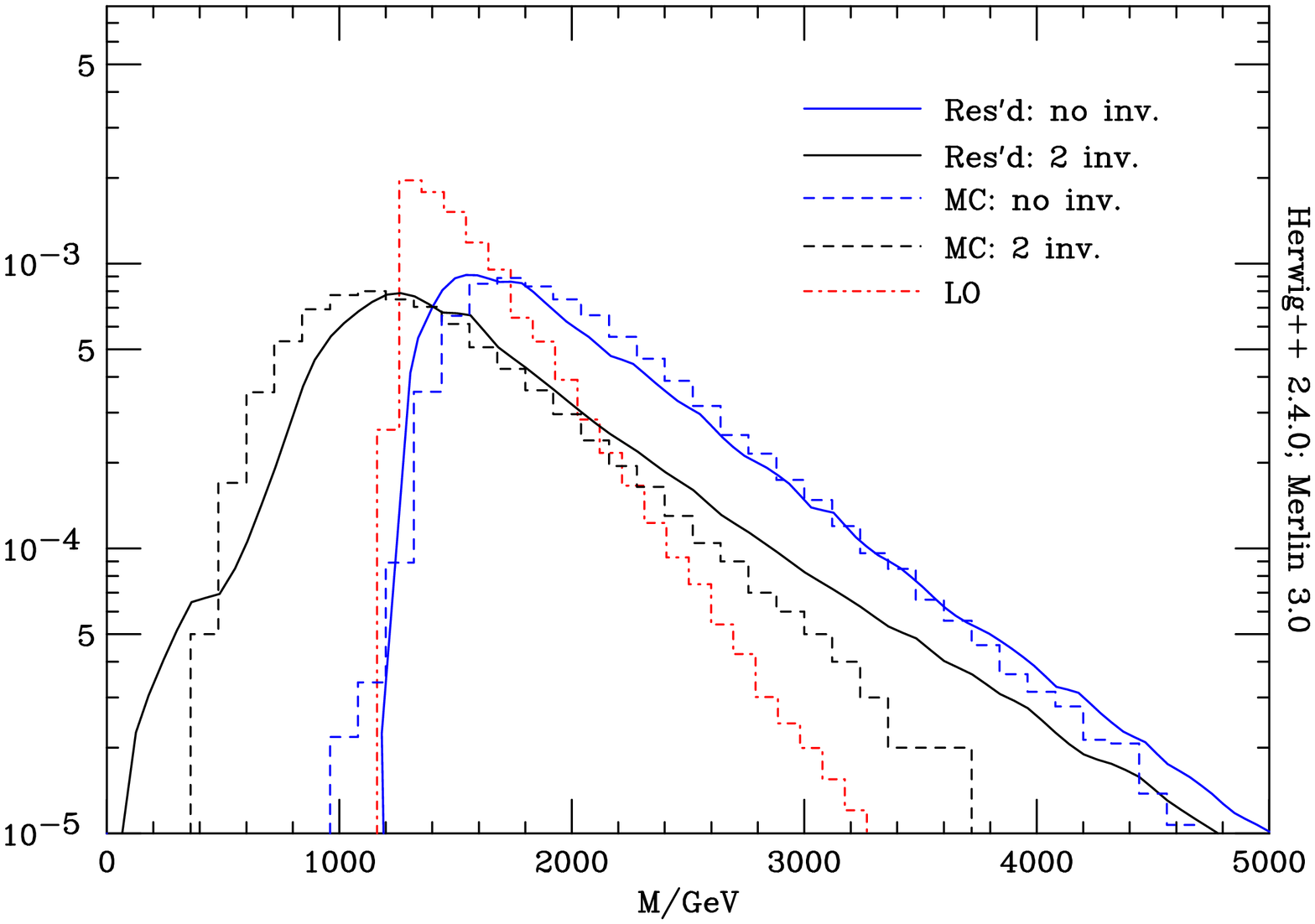}
\caption{The SPS1a gluino pair-production visible mass distributions for pseudorapidity cuts $\etam=5$ (left) and $\etam=3$ (right). The leading order distribution is shown (red dot dashes) for comparison.}
\label{fig:gge53}
\end{figure}

\begin{figure}
  \centering 
  \vspace{1.2cm}
  \includegraphics[scale=0.34, angle=90]{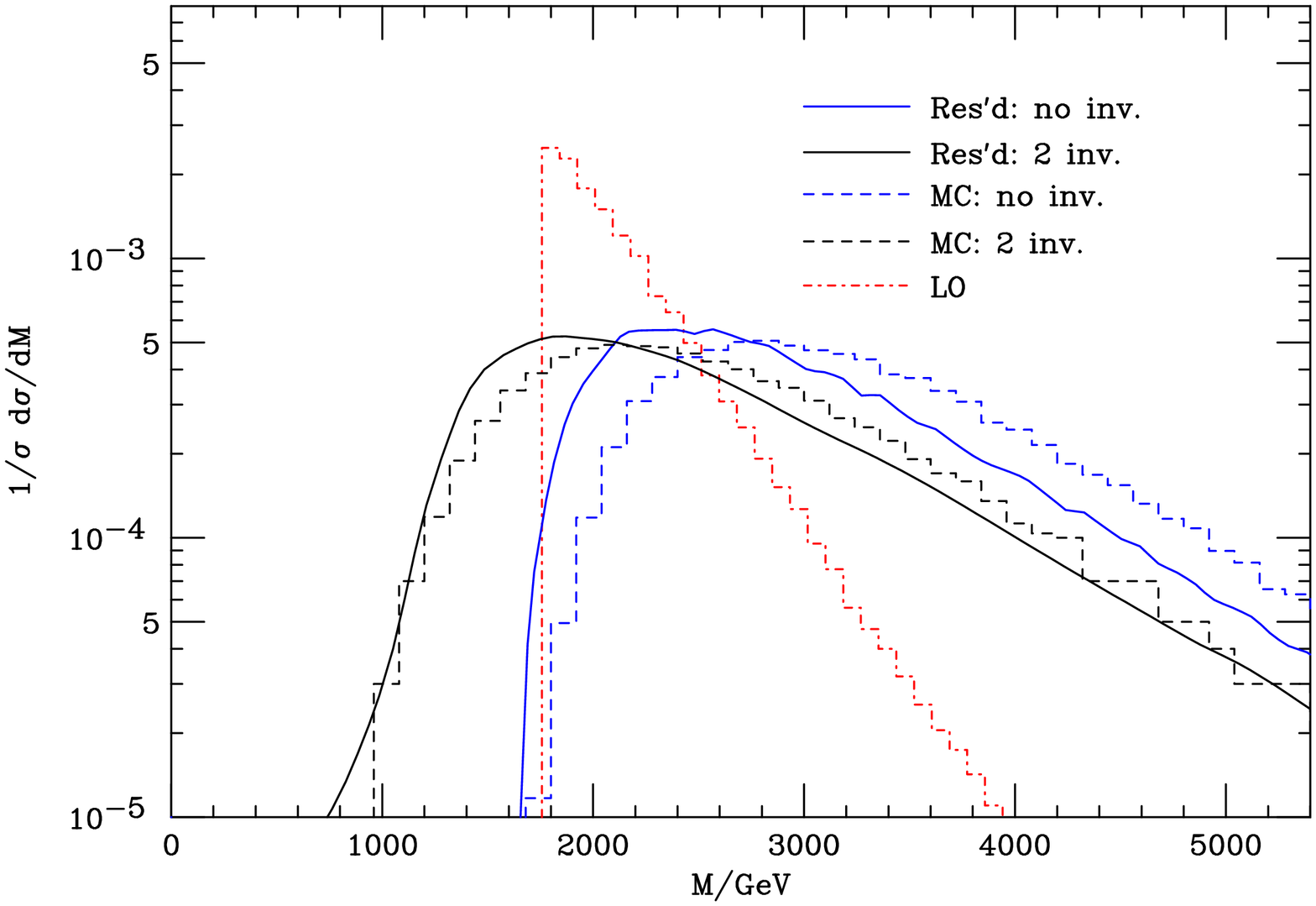}
    \hspace{0.6cm}
    \includegraphics[scale=0.34, angle=90]{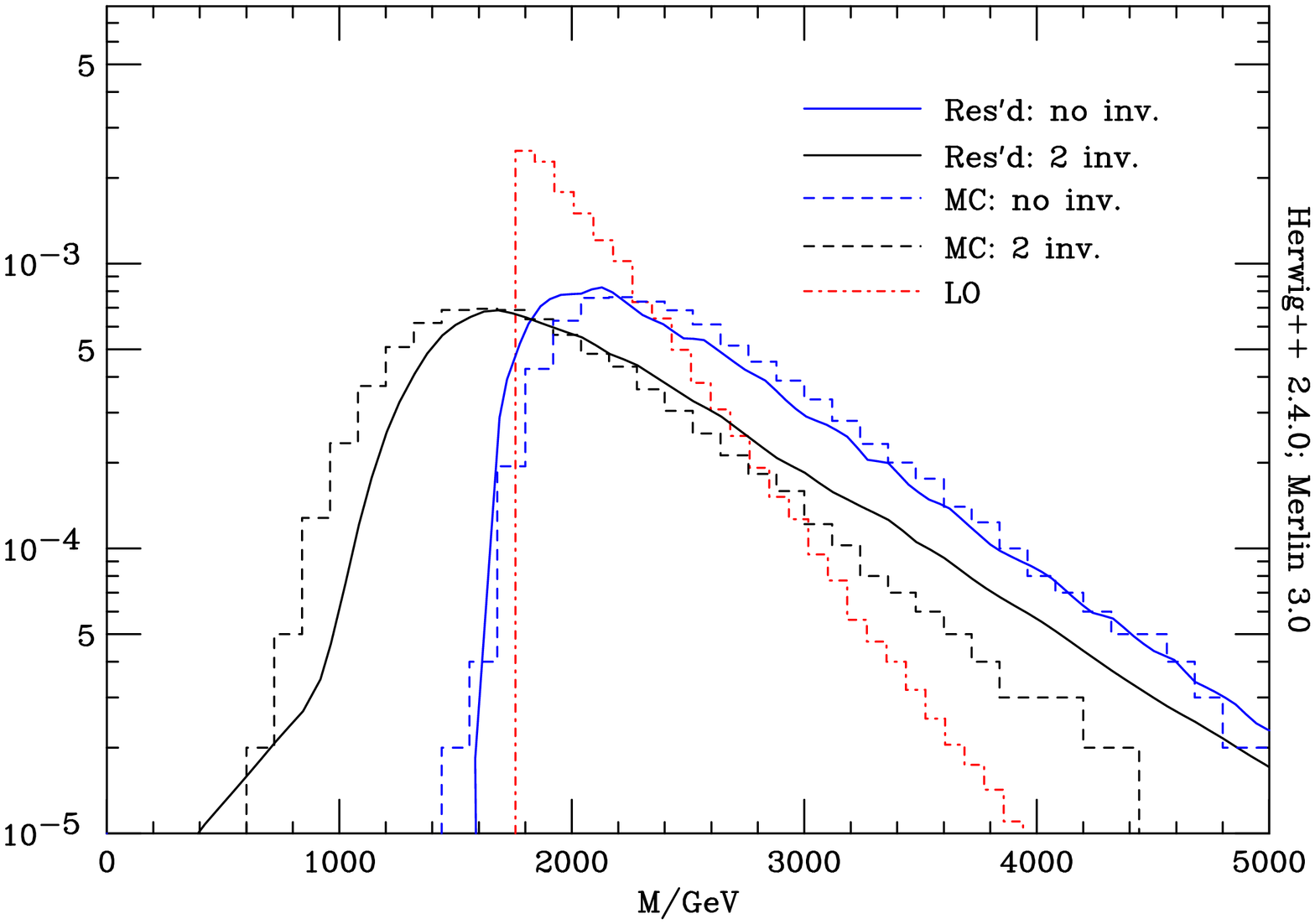}
\caption{The modified SPS1a gluino pair-production (with $m_{\tilde{g}} = 800 \gev$) results for pseudorapity cuts $\etam=5$ (left) and $\etam=3$ (right) . The leading order distribution is shown (red) for comparison.}
\label{fig:gg800e53}
\end{figure}

We focus on the SPS1a point~\cite{Allanach:2002nj}, which
has gluino and lightest neutralino masses $m_ {\tilde{g}} = 604.5
\gev$ and $m _{\chi _1 ^0 } = 97.0 \gev$ respectively (and see
table~\ref{tb:masses} for the squark masses). For simplicity we set
the squark mass in the invisible particle treatment to $550\gev$. We
also present results for a modified SPS1a point, with $m_ {\tilde{g}}
= 800 \gev$.  In this process only the two-invisibles case is
realistic, but for comparison we also show results for no invisibles,
i.e. imagining that the two lightest neutralinos are also detected.
\begin{table}[htb]
\begin{center}
\begin{tabular}
{|c|c|c|c|} \hline
Particle& Mass (GeV)& Particle& Mass (GeV) \\ \hline
$\tilde{g}$&604.5&$\tilde{s}_L$&570.7 \\ \hline
$\chi _1 ^0$& 97.0&$\tilde{s}_R$&547.9 \\ \hline
$\tilde{u}_L$&562.3&$\tilde{b}_1$&515.3 \\ \hline
$\tilde{u}_R$&548.2&$\tilde{b}_2$&547.7 \\ \hline
$\tilde{d}_L$&570.7&$\tilde{t}_1$&400.7 \\ \hline
$\tilde{d}_R$&547.9&$\tilde{t}_2$&586.3 \\ \hline
\end{tabular}
\end{center}
\caption{The relevant particle masses in the supersymmetric model used
  in the invisible study, SPS1a. The modified SPS1a point differs in that it has $m_{\tilde{g}} = 800 \gev$.}
\label{tb:masses}
\end{table}
When $\etam = 5,3$, there is fairly good agreement between the Monte Carlo and resummation
predictions in both the two-invisibles and no-invisibles cases, and
for both gluino masses, as shown in Figs.~\ref{fig:gge53}
and~\ref{fig:gg800e53}, where one should compare the dashed histograms
(Monte Carlo) to the solid curves of the same colour (resummation).

The shift in the peak of the visible mass distribution in going from
no  to two invisibles is much larger than that in top pair production,
amounting to 600-700 GeV, roughly independent of $\etam$ and the
gluino mass.  This results mainly from the higher masses of the
intermediate particles in the decays ($m_{\tilde q}\simeq 550$ GeV
vs. $m_W=80$ GeV), which implies a higher energy release, rather than
the masses of the invisible particles themselves ($m_{\chi_1^0} = 97$
GeV vs. $m_\nu=0$).

\begin{figure}
  \centering 
  \vspace{1.5cm}
    \includegraphics[scale=0.35, angle=90]{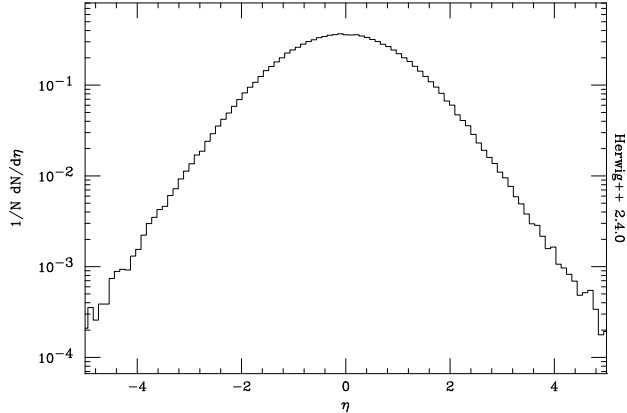}
\caption{The SPS1a gluino pair-production pseudorapidity distribution for $m_{\tilde{g}} = 604.5 \gev$.}
\label{fig:gprodrap}
\end{figure}

\begin{figure}
  \centering 
  \vspace{1.2cm}
    \includegraphics[scale=0.34, angle=90]{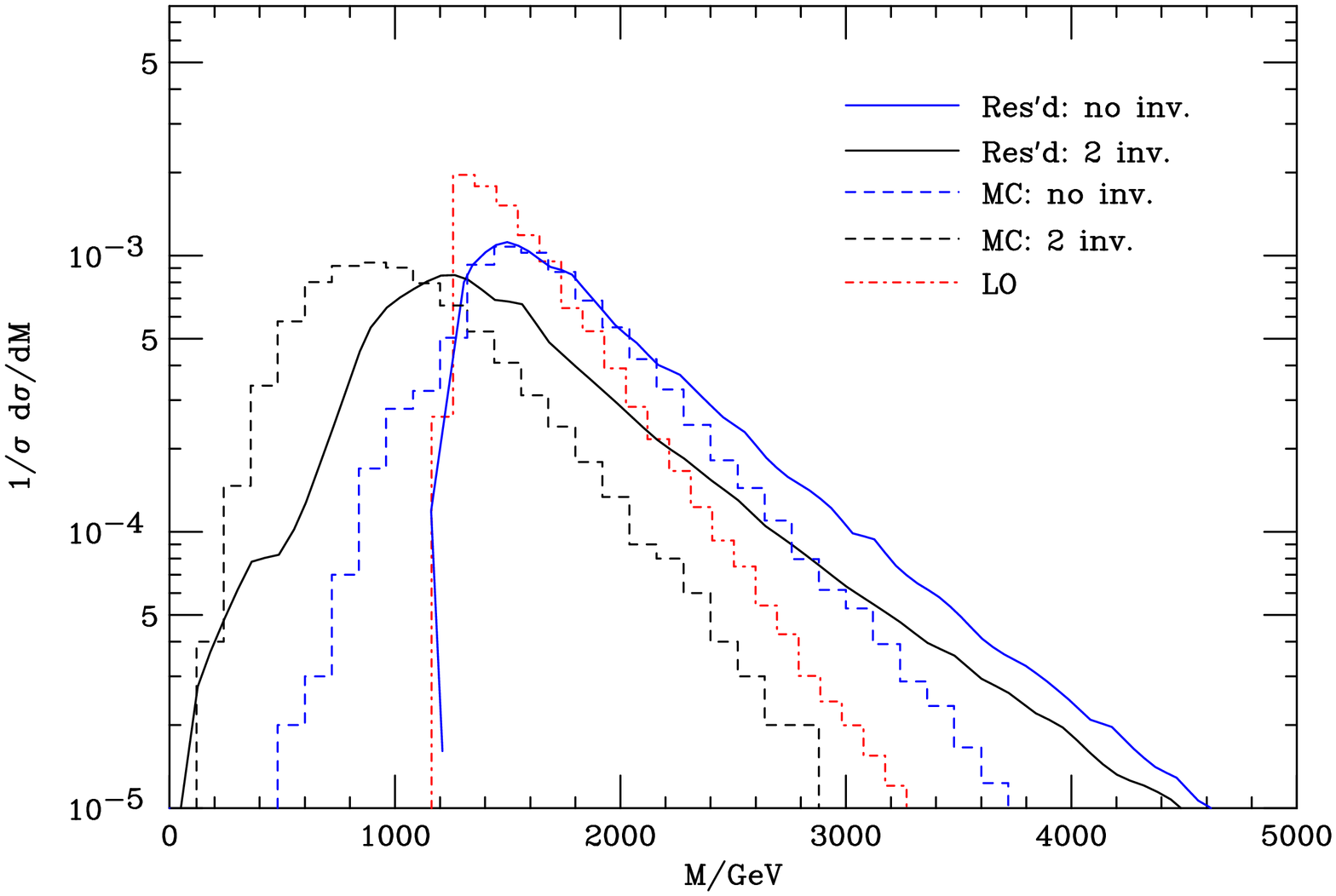}
    \hspace{0.5cm}
    \includegraphics[scale=0.34, angle=90]{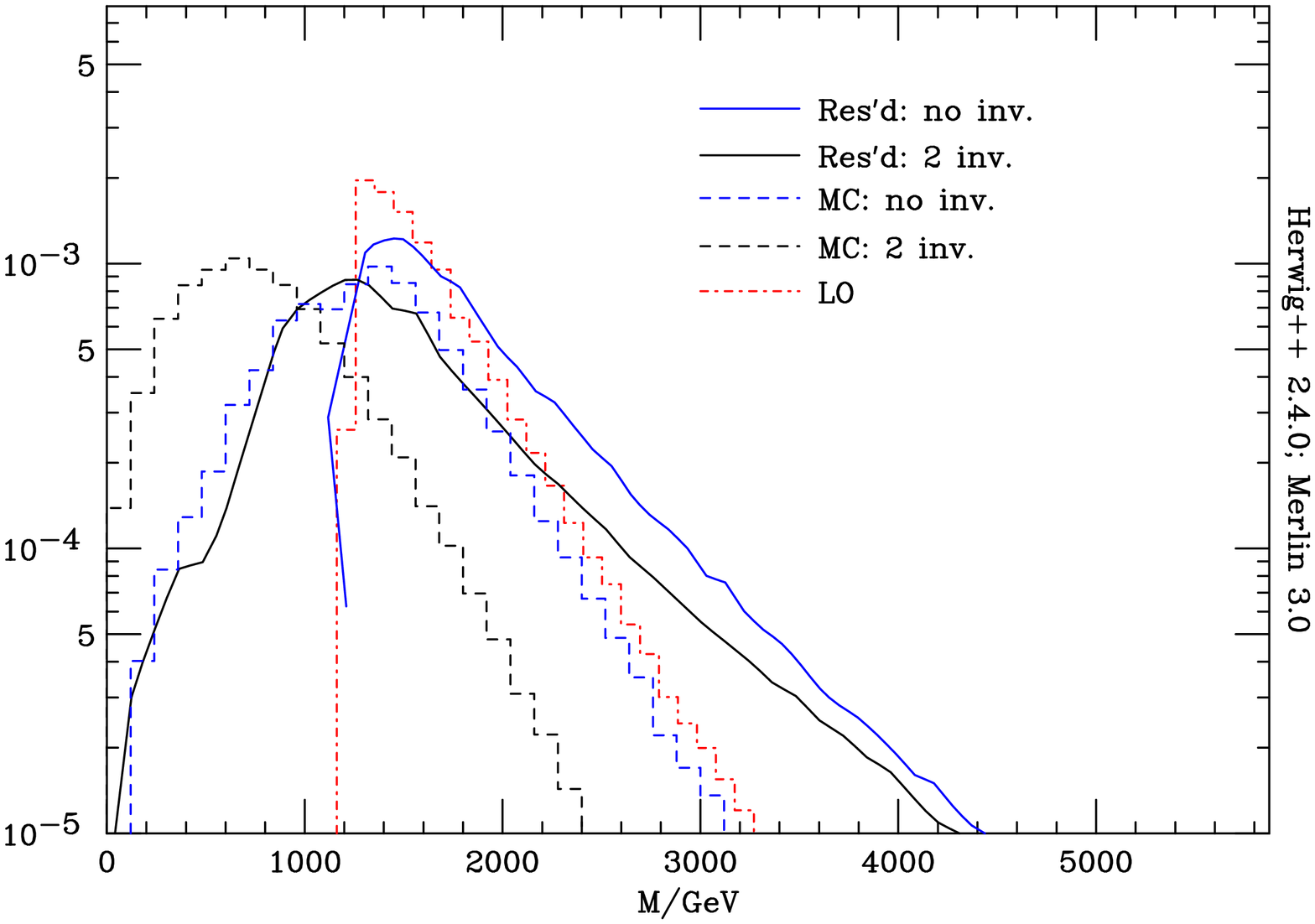}
\caption{The SPS1a gluino pair-production results for pseudorapidity cuts $\etam=2$ (left) and $\etam=1.4$ (right). The leading order distribution is shown (red dot dashes) for comparison.}
\label{fig:gge214}
\end{figure}

\begin{figure}
  \centering 
  \vspace{1.2cm}
    \includegraphics[scale=0.34, angle=90]{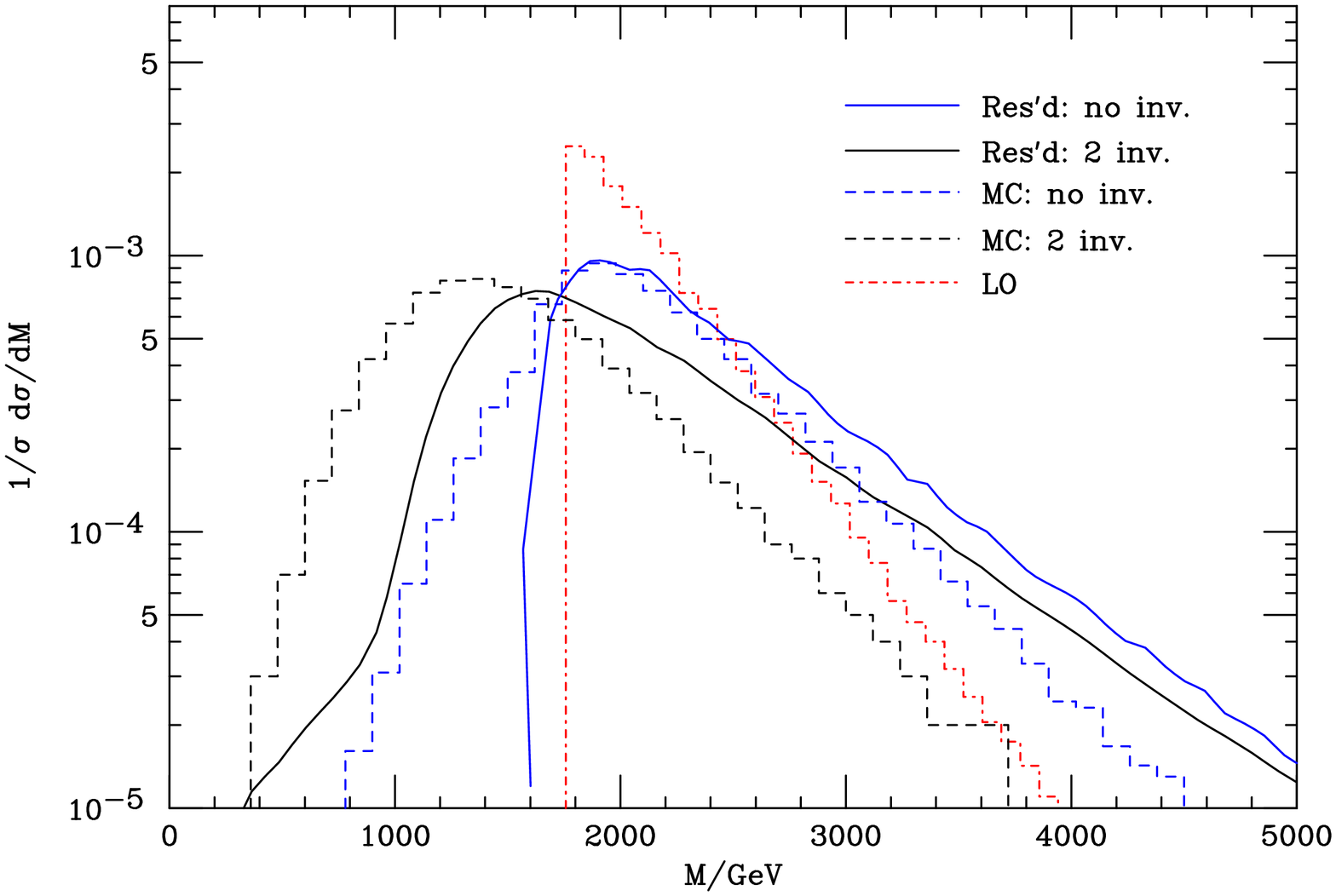}
    \hspace{0.5cm}
    \includegraphics[scale=0.34, angle=90]{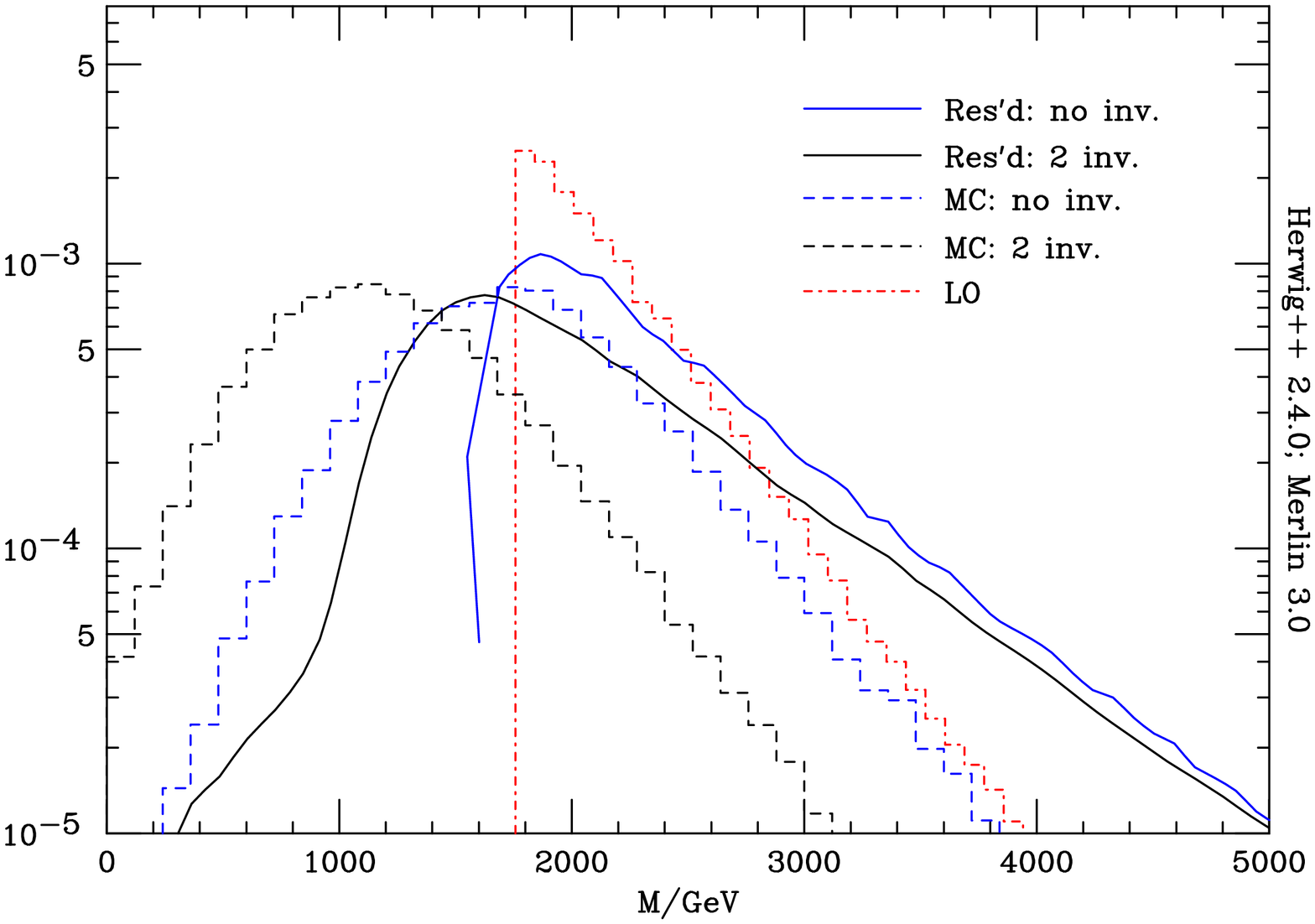}
\caption{The modified SPS1a gluino pair-production (with $m_{\tilde{g}} = 800 \gev$) results for pseudorapity cuts $\etam=2$ (left) and $\etam=1.4$ (right). The leading order distribution is shown (red) for comparison.}
\label{fig:gg800e2}
\end{figure}

One of the assumptions of the resummation is that all the visible hard
process decay products are detected, which is not true when the
maximum pseudorapidity $\etam$ is restricted to lower
values. When $\etam \sim 2$ in the Monte Carlo analysis, a significant
number of hard process particles begin to be excluded and hence the
curves shift to lower values than the resummed
predictions. Figure~\ref{fig:gprodrap} shows the rapidity distribution
of the decay products of the gluino at parton level for $m_
{\tilde{g}} = 604.5 \gev$. For the case shown, cuts of $\etam = 5,3,2$
and $1.4$ correspond to exclusion of, respectively, $\sim$0.002\%,
1.1\%, 7.5\% and 20.0\% of the gluino decay products from the
detector. The effect of this appears in Figs.~\ref{fig:gge214}
and~\ref{fig:gg800e2}, where the Monte Carlo distributions are
narrower and peak at lower masses than the resummed predictions.  The
variation between the resummed $\etam = 2$ and 1.4 curves is
smaller than that between $\etam = 5$ and 3,
since they correspond to smaller differences in $Q_c$.

The heavy and light gluino scenarios exhibit similar behaviour when
varying the pseudorapidity coverage and the number of invisibles,
showing the lack of dependence of the resummation on the mass of the
pair-produced particle. The sensitivity to low-scale PDF behaviour and
showering is reduced compared to the $t\bar{t}$ case since we are
considering higher centre-of-mass energies, with the lowest possible
$Q_c$ now being of the the order $2 m_{\tilde{g}} \times e^{-5} \sim 8
\gev$. The position of the curves is also sensitive to the precise
definition of $Q_c$ in terms of $\etam$.

\begin{table}[h!]
\begin{center}
\begin{tabular}
{|c|c|c|c|} \hline
$m_{\tilde{g}}$ (GeV).& $\eta_{\mathrm{max}}$& MC (GeV) (0 inv./2 inv.) & Resum. (GeV) (0 inv./2 inv.) \\ \hline
604.5&5& 2280/1560 & 1785/1620 \\ \hline
604.5&3& 1680/1080& 1593/1204 \\ \hline
604.5&2& 1440/840& 1497/1204\\ \hline
604.5&1.4& 1380/660 & 1497/1204 \\ \hline
800.0&5& 2820/2100& 2569/1870 \\ \hline
800.0&3& 2220/1620& 2128/1684 \\ \hline
800.0&2& 1920/1380& 1865/1683 \\ \hline
800.0&1.4& 1740/1140& 1865/1683 \\ \hline
\end{tabular}
\end{center}
\caption{Summary of the positions of the
  peaks of the gluino pair-production visible mass distributions as
  given by the Monte Carlo and the resummation, for different values
  of the maximum pseudorapidity and for no and two invisibles.}
\label{tb:peaks}
\end{table}

Table~\ref{tb:peaks} shows a summary of the peak
positions for all cases and different pseudorapidity cuts.  For the
higher values of $\etam$, the agreement between the Monte Carlo and
resummation is satisfactory.  There is a large difference in the peak
positions for no invisibles and $\etam=5$, but this is
mainly due to the broad shape of the peak in this case, while the
overall distributions agree better.  For $\etam\leq 2$ there is a
growing discrepancy, especially for the realistic case of two
invisibles, due to the loss of particles coming from the hard process.

\subsection{\bm Hadronization effects}\label{sec:hadronization}
We have assumed that ISR partons emitted at pseudorapidities above
$\etam$ do not contribute to the visible invariant mass.  This would
be true if the hadronization process were perfectly local in
angle. However, as a result of  hadronization  high-rapidity ISR
partons can produce lower-rapidity hadrons and thus `contaminate' the
detector and shift the visible mass to higher values.

\begin{figure}[htb]
  \centering 
  \vspace{1.0cm}
    \includegraphics[scale=0.30, angle=90]{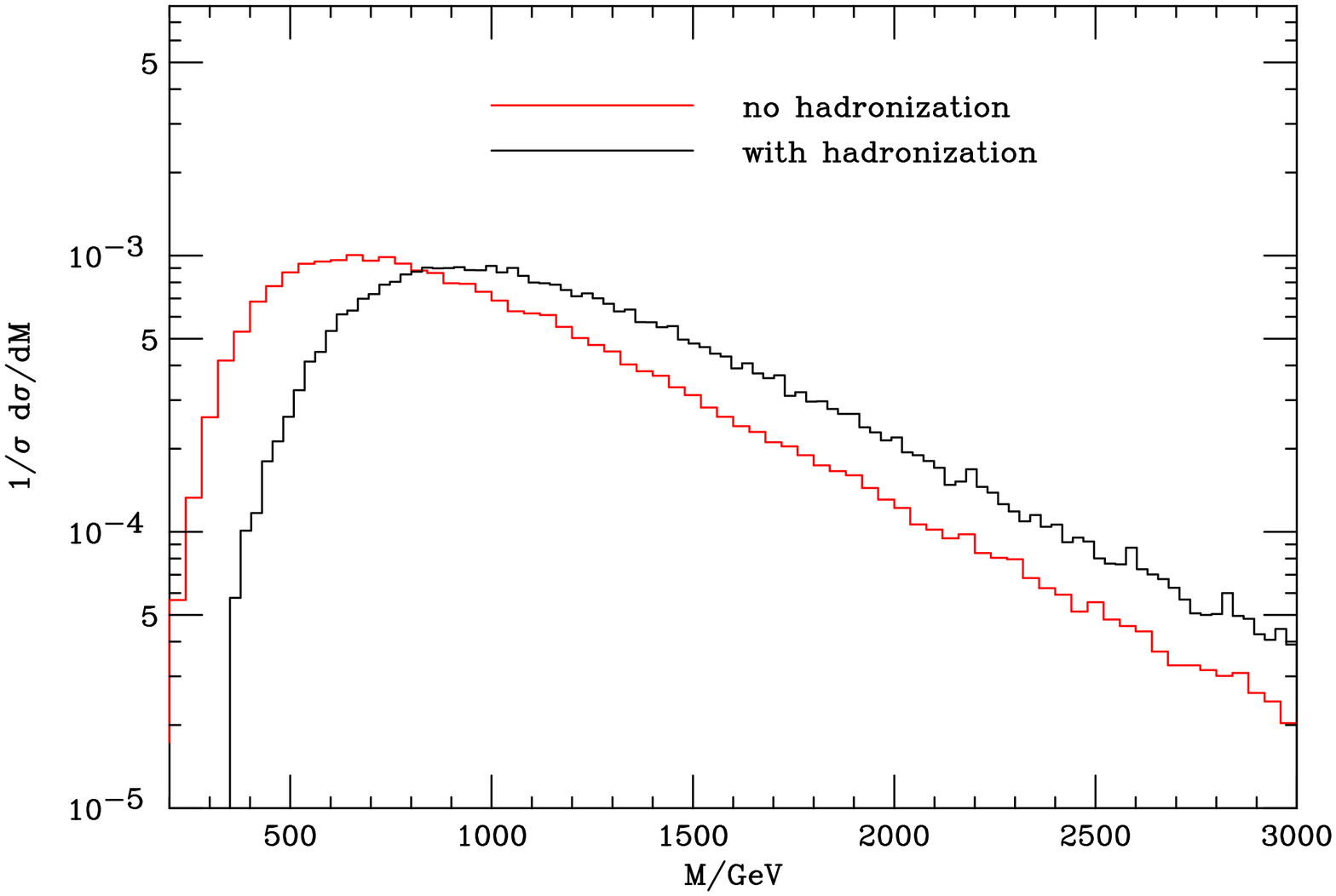}
   \hspace{1.0cm}
    \includegraphics[scale=0.30, angle=90]{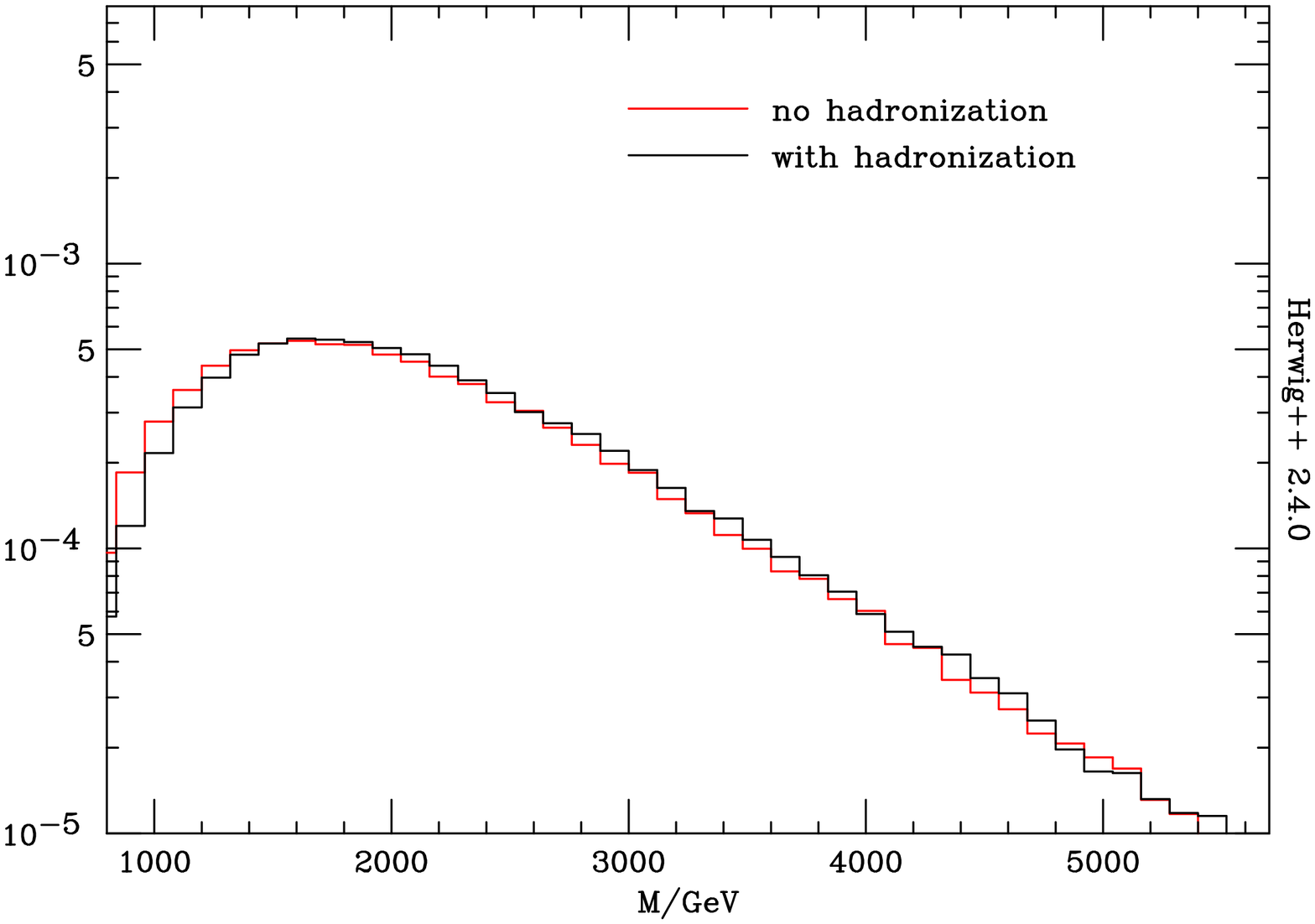}
    \caption{The $t\bar{t}$ fully semi-leptonic (left) and SPS1a gluino pair-production (right, with $m_{\tilde{g}} = 604.5 \gev$) visible mass distributions for a pseudorapity cut $\etam=5$ with and without hadronization (black and red respectively).}
\label{fig:hadroniz}
\end{figure}

The hadronization model employed in the \texttt{Herwig++} Monte Carlo
is a refinement of the cluster model described in
ref.~\cite{Webber:1983if}. The model involves clustering of partons
into colour-singlet objects that decay into hadrons, resulting in a smearing of the
pseudorapidity distribution which causes the increase in the visible
mass described above.  The effect  is shown in Fig.~\ref{fig:hadroniz}
for gluino and top pair-production (excluding the invisible particles
from the hard process). The effect was found to be larger for
$t\bar{t}$ production where the mass distribution is shifted
significantly, whereas in gluino pair production the shift is negligible.

\subsection{Underlying event}\label{sec:MPI}
\begin{figure}[htb]
  \centering 
  \vspace{1.0cm}
    \includegraphics[scale=0.30, angle=90]{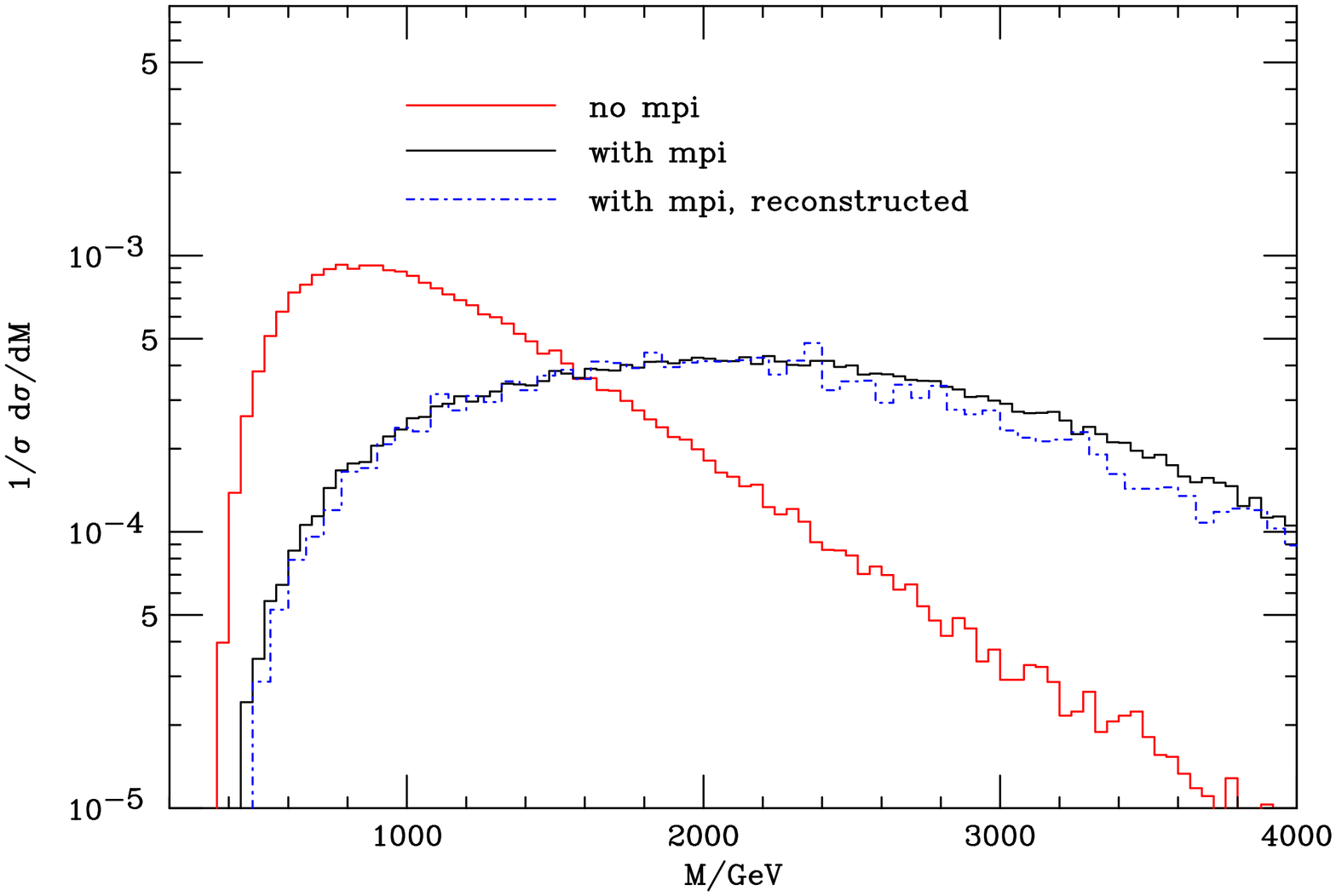}
    \hspace{1.0cm}
    \includegraphics[scale=0.30, angle=90]{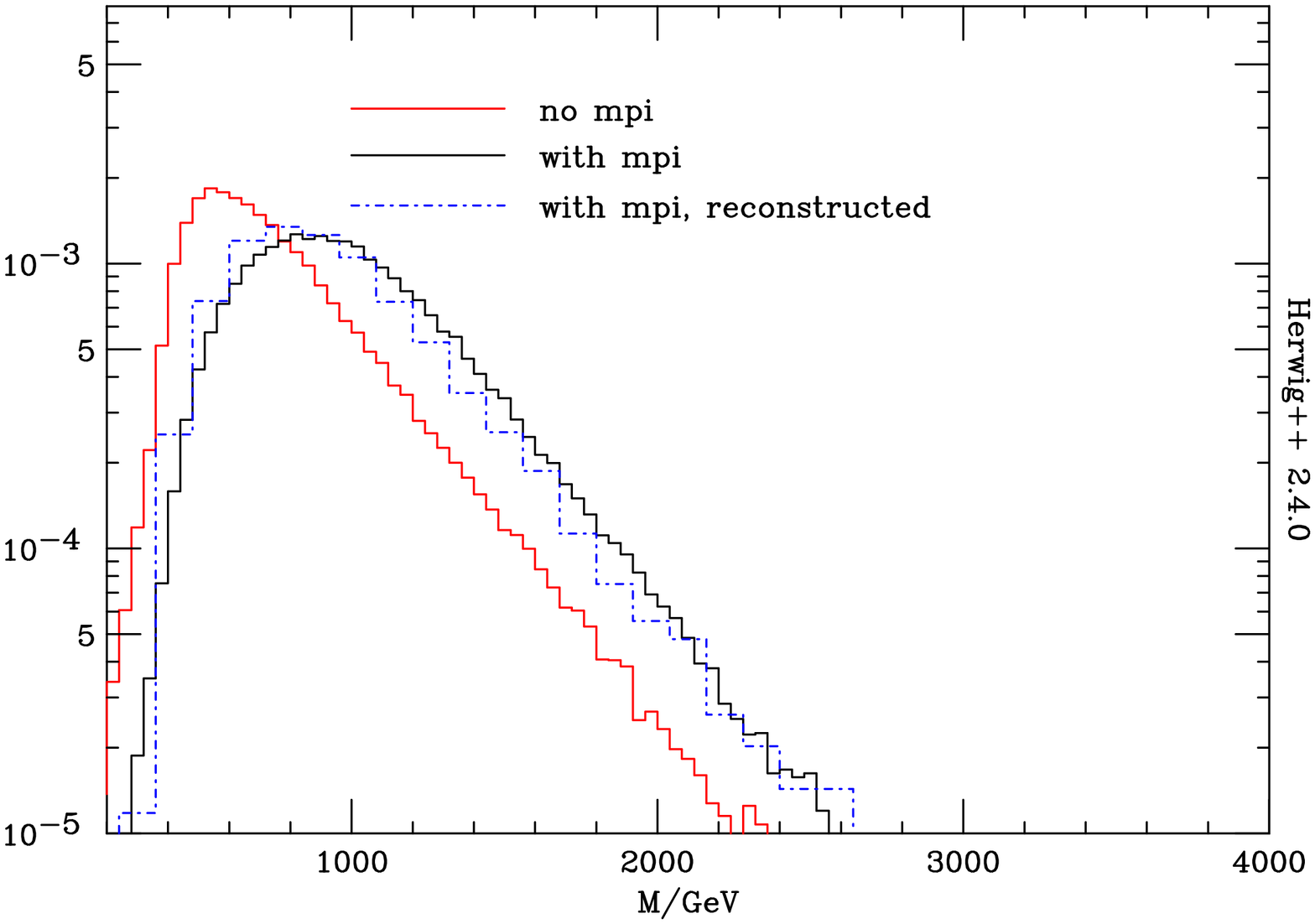}
    \caption{The $t\bar{t}$ fully hadronic visible mass
      distributions for pseudorapity cuts $\eta_{max}=5$ (left) and
      $\eta_{max}=3$ (right),  with and without multiple parton
      interactions (black and red respectively) and the reconstructed
      curves (blue dot dashes). The $\etam = 5$ curve was
      reconstructed using the resummed results for the visible mass
      and rapidity, whereas the $\etam = 3$ curve was reconstructed
      using the Monte Carlo visible mass and rapidity.}
\label{fig:mpi_tt}

\end{figure}
\begin{figure}[htb]
  \centering 
  \vspace{1.0cm}
    \includegraphics[scale=0.30, angle=90]{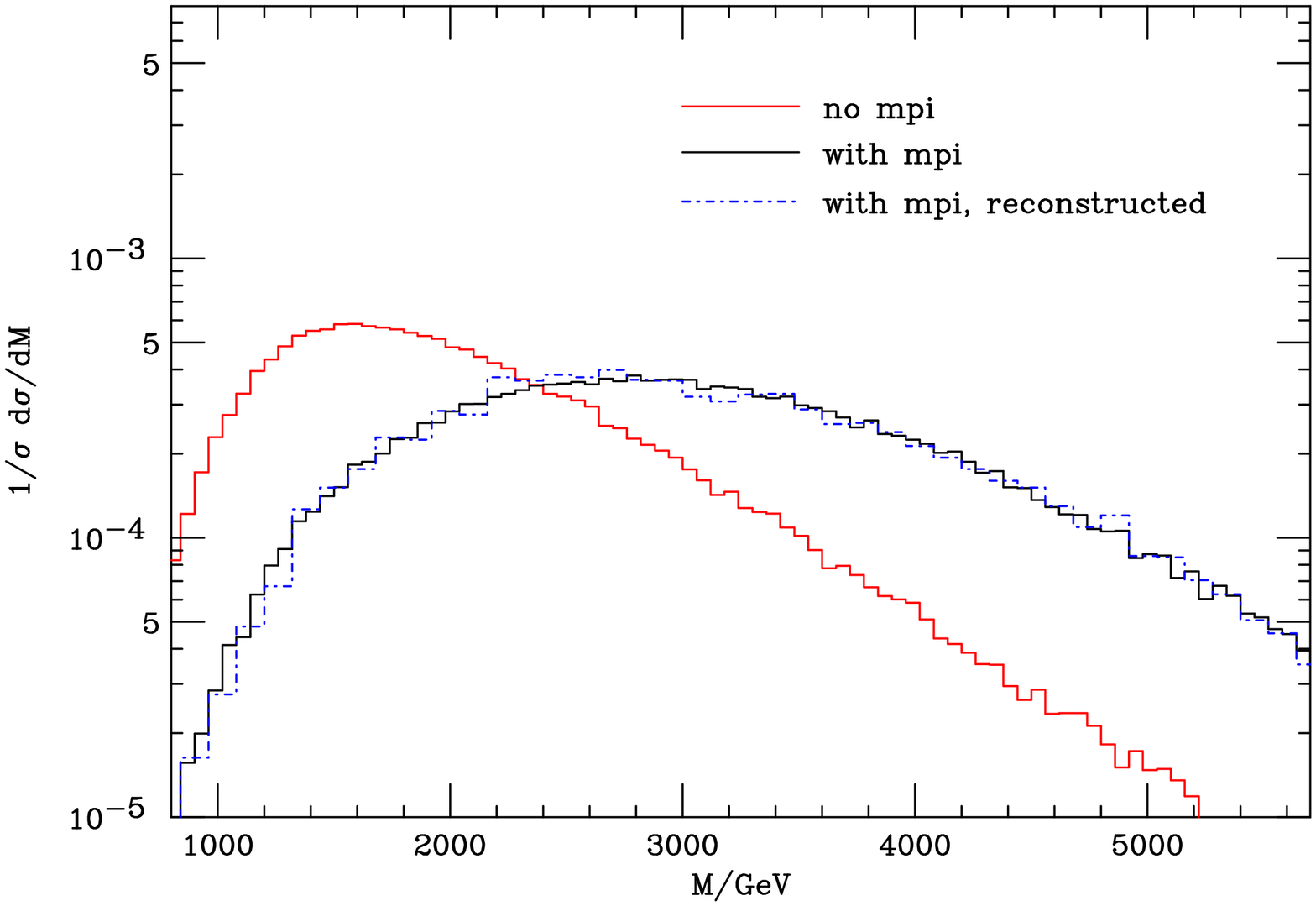}
    \hspace{1.0cm}
    \includegraphics[scale=0.30, angle=90]{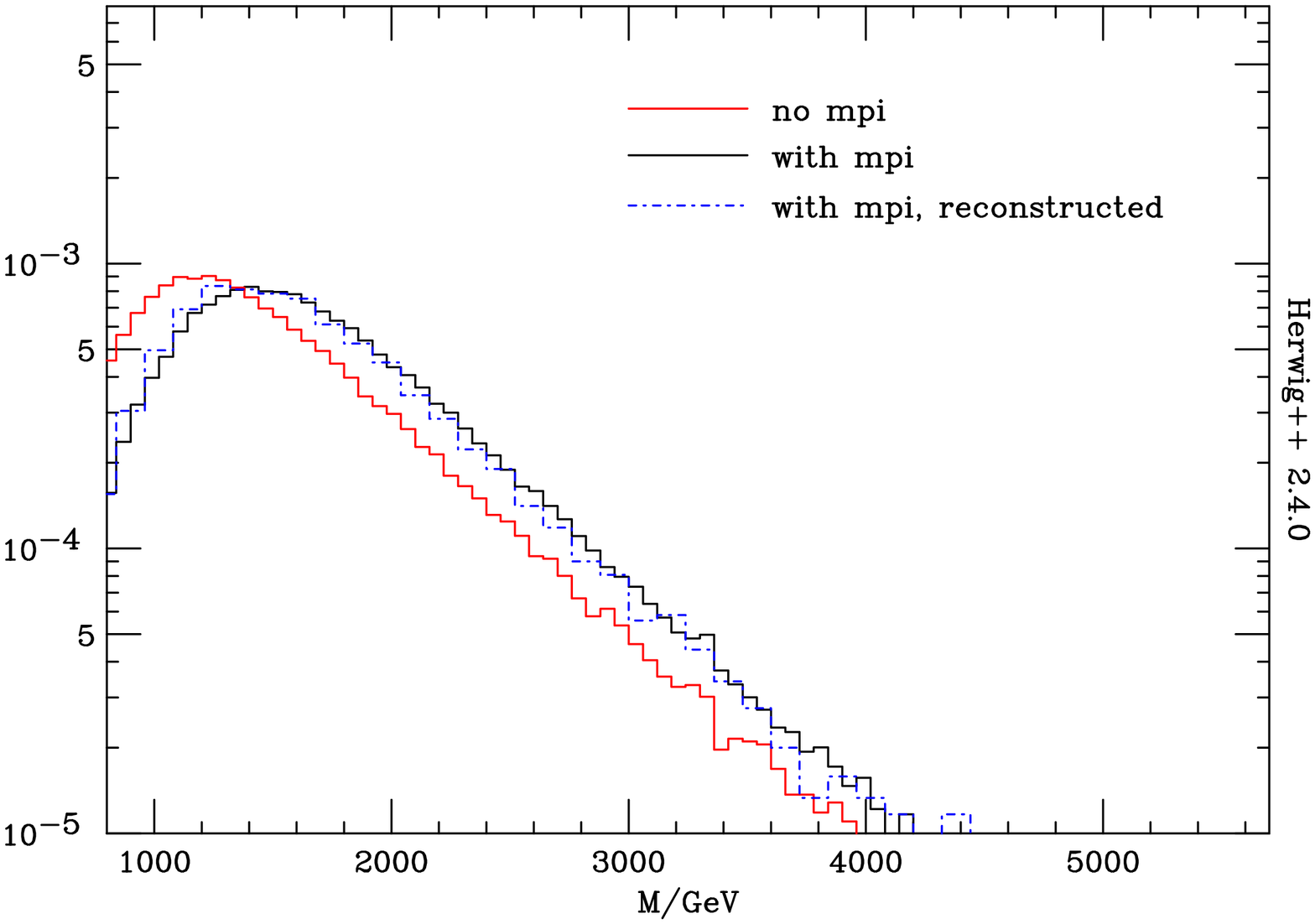}
    \caption{The SPS1a gluino pair-production (with $m_{\tilde{g}} =
      604.5 \gev$) visible mass distributions for pseudorapity cuts
      $\eta_{max}=5$ (left) and $\eta_{max}=3$ (right),  with and
      without multiple parton interactions (black and red
      respectively) and the reconstructed curves from the Monte Carlo visible masses and rapidities (blue dot dashes).}
\label{fig:mpi_gg}
\end{figure}

The underlying event, which is thought to arise from multiple
soft interactions of spectator partons, is a further source of
non-perturbative contributions to the visible mass.  If $P_H^\mu$
represents the ``hard'' visible 4-momentum studied in earlier sections
and $P_U^\mu$ represents that due to the underlying event, the total
visible mass is given by
\bea\label{eq:Mtot}
M^2 = (P_H+P_U)^2 &=& M_H^2 + M_U^2 +2 (E_HE_U-P_{zH}
P_{zU})\nonumber\\
&=& M_H^2 + M_U^2 +2 M_U\sqrt{M_H^2+\met^2} \cosh(Y_H-Y_U)\;.
\eea
where we neglect transverse momentum associated with the underlying
event. Thus, even if the visible invariant mass due to the underlying event
is small, its effect on the overall visible mass may be enhanced through
the last term on the right-hand side.

The underlying event is simulated in \texttt{Herwig++}  by a multiple
parton interaction model along the lines of
ref.~\cite{Butterworth:1996zw}. In this model, for the rapidity
ranges considered here, the underlying event is approximately
process-independent and exhibits little correlation
with the rest of the event.  Therefore, to a good approximation, the
distributions of the variables related to the underlying event, $Y_U$
and $M_U$, can be determined once and for all at each collider energy. The
process-dependence comes primarily through the dependence on $Y_H$ and
$M_H$, which can be calculated using the resummation formula given in
eq.~(\ref{eq:SigmaDefinv}). The overall visible mass distribution can
then be obtained by convolution using eq.~(\ref{eq:Mtot}).

The effects of including the underlying event in the visible mass
distribution are shown in Figs.~\ref{fig:mpi_tt} and \ref{fig:mpi_gg}
for $t\bar t$ and gluino pair production, respectively. The multiple
parton interactions push the peak value to substantially higher
masses.  The shift amounts to about 250 GeV at $\eta_{max}=3$ and 1.2
TeV at $\eta_{max}=5$, and is roughly process-independent.  However,
since the underlying event is approximately uncorrelated with the hard
process, the visible mass distributions can be reconstructed well by
the convolution procedure outlined above, as shown by the blue
dot-dashed curves in Figs.~\ref{fig:mpi_tt} and \ref{fig:mpi_gg}.
These features of the underlying event will need to be validated by
LHC data on a variety of processes.  Accurate modelling of the
underlying event is important for practically all aspects of hadron
collider physics.

\section{Conclusions}\label{sec:conc}
In this paper we have presented detailed predictions on the total
invariant mass $M$ of the final-state particles registered in a detector,
as a function of its pseudorapidity coverage $\etam$.  This quantity
provides the dominant contribution to many global inclusive observables
such as the new variable $\hat{s}^{1/2}_{\rm min}$ in (\ref{eq:smin_def}), which can provide
information on the energy scales of hard processes.
We have extended the resummation method presented
in~\cite{Papaefstathiou:2009hp} to include the effects of invisible
particle emission from the hard process.
We have considered the case of one or two invisible
particles and presented results for Standard Model top quark pair
production and SPS1a gluino pair production, obtained using a
numerical Mellin moment inversion method.

In the case of $t\bar{t}$ production the invisible particles are
neutrinos from W-boson decays and their effect on the visible
invariant mass distribution is small, even when both decays are
leptonic.  This is mainly a consequence of the small W-boson mass
compared to the overall invariant mass, rather than the negligible
neutrino mass. For gluino pair production the invisibles are a pair of
massive LSPs  from squark decays.  The LSP mass is again small
compared to the overall invariant mass, but the squark masses are not,
leading to a substantial downward shift in the visible mass
distribution, of the order of the squark mass. In both cases the
resummed predictions are in fair agreement with Monte Carlo estimates
of the position of the peak in the distribution, provided the
pseudorapidity range covered by the detector is large enough
($\etam\gtap 3$). For $\etam\sim 3$, the difference between the Monte
Carlo prediction and resummed predictions is of the order of 100 GeV
for both the heavy and light gluino SPS1a points. The agreement
becomes worse when the pseudorapidity range is restricted, due to
particle loss from the hard process. Table~\ref{tb:peaks} shows the
positions of the peaks of the distributions for the Monte Carlo
results from \texttt{Herwig++} and the resummation.
 
These comparisons were made with
Monte Carlo visible mass distributions at parton
level. We found that non-perturbative effects,
especially the underlying event, tend to shift the invariant mass
distributions to significantly higher values than expected from a
purely perturbative calculation.
 According to the underlying event model used in \texttt{Herwig++},
 the shift amounts to about 250 GeV at $\eta_{max}=3$ and 1.2 TeV at $\eta_{max}=5$.
This effect is also expected in other observables sensitive to longitudinal momentum
components, such as $\hat{s}^{1/2}_{\rm min}$.  However, in this model
the underlying event is only weakly correlated with the rest of the
event and hence its effects can be determined once and for all at each
collider energy. The modelling of the underlying event is an
important feature of the Monte Carlos that needs to be validated by
comparison with experiment.  Once this has been done, a wide range of
global inclusive observables, including the visible invariant mass,
will be reliably predicted and useful for establishing the scales of
contributing hard subprocesses.

\section*{Acknowledgements}
BW is grateful to the CERN Theory Group and the Aspen Center for
Physics for hospitality during parts of this work, which was supported
in part by the UK Science and Technology Facilities Council and the
European Union Marie Curie Research Training Network MCnet
(contract MRTN-CT-2006-035606).

\appendix
\section{Calculation of the evolution kernels}\label{app:mellin}
The evolution kernels, $K_{a'a}(z)$, were calculated using the
Mellin inverse transform method.  Recall that the Mellin transform is
defined by the expression
\beq
K^{b'b}_N = \int_0^1 dz\,z^{N-1}K_{b'b}(z)
\eeq
with inverse
\begin{equation}\label{eq:mellinK}
K_{a'a}(z) = \frac{1}{2 \pi i} \int_C \mathrm{d} N z^{-N} K_N^{a'a}\;,
\end{equation}
where the contour $C$ runs parallel to the imaginary axis and is to the right of all
singularities of the integrand.

Taking into account the running of the strong coupling, the Mellin
transform of the solution of the evolution equation (\ref{eq:evolK}) is
\beq\label{eq:Kba}
 K^{ba}_N = \left(\left[\frac{\alps(Q_c)}{\alps(Q)}\right]^{p\Delta_N}\right)_{ba}
\eeq
with $p = 6/(11C_A-2n_f)$ and
\beq
(\Delta_N)_{ba} = \frac\pi{\alps}(\Gamma_N)_{ba} = \int_0^1 dz\,z^{N-1}P_{ba}(z) \ .
\eeq
When computing this expression we need to take into account the fact
that $\Delta_N$ is a matrix. This can be done most easily by writing
it as
\begin{equation}\label{eq:matrixexponential}
K^{a'a}_N = \left(\mathcal{O}^{-1} \left[  \frac{\alpha_S(Q_c)}{\alpha_S(Q)} \right]^{p\,\mathrm{diag}(\lambda_{N,i})} \mathcal{O} \right)_{a'a}\;,
\end{equation}
where $\mathcal{O}$ is the matrix of eigenvectors of $\Delta_N$ and
$\mathrm{diag}(\lambda_N,i)$ is the diagonal matrix of its
eigenvalues. This is equivalent to using, implicitly, the singlet and
non-singlet basis for the PDF
evolution~\cite{qcdcollider}.

\begin{figure}[htb]
  \begin{center}
    \includegraphics[scale=0.50, angle=90]{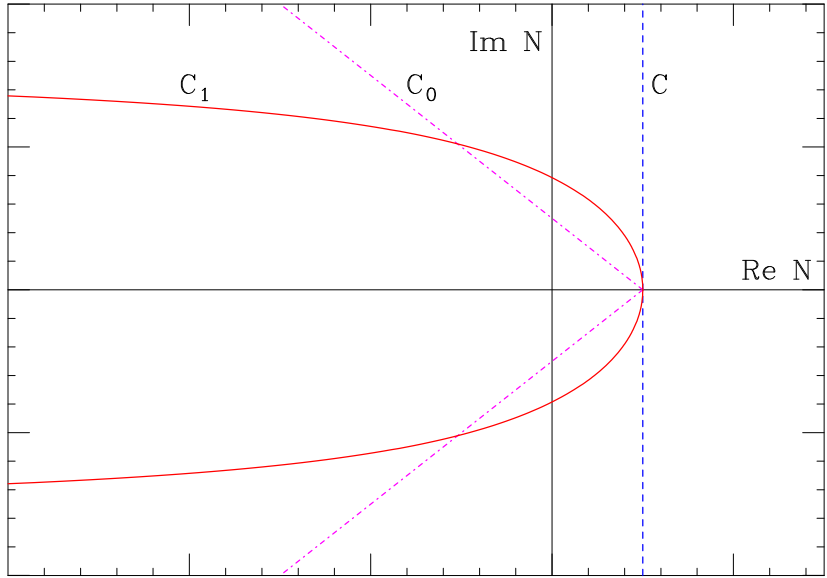}
  \end{center}
\caption[]{Integration contours, $C_0$ and
  $C_1$, for the inverse Mellin transform as given by the
  Bromwich integral, eq.~(\ref{eq:mellinK}).}

\label{fig:contours}
\end{figure}

To evaluate the inverse transform (\ref{eq:mellinK}), numerical
convergence of an otherwise infinitely oscillatory expression is
achieved by choosing a contour that introduces a damping factor.
A method commonly used in PDF evolution (see, for example,~\cite{Vogt:2004ns}) is
to rotate the upper and lower portions of the vertical contour
so that they slope back into the left half-plane, as shown
by $C_0$ in Fig.~\ref{fig:contours}.  This introduces exponential
damping along the contour, which is chosen so as to
enclose the same singularities as the `Bromwich' contour
$C$, and therefore converges to the correct result by Cauchy's theorem.
However, in the case of the evolution kernels $K_{a'a}(z)$ the
linear contour $C_0$ does not  provide sufficient accuracy to reproduce the
function from its transform. This is due to its inability to invert
the constant function $f_N = c$ to the correct analytic result, a
delta function. This implies that the inversion does not
reproduce the necessary initial condition of the evolution,
$K_{a'a}(z, Q=Q_c) \varpropto \delta(1-z)$. A numerically more
accurate contour is available in the literature, used in the so-called
`Fixed-Talbot algorithm'. This contour $C_1$ has the form $\mathrm{Re}(N) =
\mathrm{Im} N \cot ( \mathrm{Im} N / r )$ where $r$ is a parameter
which we set to $r = 0.4 m/\log(1/z)$ during the computation, $m$
being the required precision in number of decimal digits, a value
derived from numerical experiments. The contour is related to the
steepest descent path for a certain class of functions. For further
details on its origin and accuracy see~\cite{Abate:2004}.

\section{Pair-production cross sections}\label{app:cross-sections}
The leading-order parton-level cross section for QCD pair-production of particles of mass $m_p$ may be written in terms of scaling functions $f_{ij}$ as
\begin{equation}
\hat{\sigma}_{ij} (Q^2) = \frac{\alpha_S^2(Q^2)}{m_p^2} f_{ij}\;.
\end{equation}

For heavy quark pair-production, the functions for gluon-gluon and quark-antiquark initial states are given by~\cite{qcdcollider}:
\begin{align}
f_{gg} =& \frac{ \pi \beta \rho } { 27 } (2 + \rho)\;,\\ 
f_{q\bar{q}} =& \frac{ \pi \beta \rho } {192 } \left[ \frac{1}{\beta} ( \rho ^2 + 16 \rho + 16 ) \log\left(\frac{1+\beta}{1 - \beta}\right) - 28 - 31 \rho \right]\;.
\end{align}
where $\rho= 4 m_q^2 / Q^2$ and $\beta = \sqrt{1 - \rho}$.
 For the case of gluino pair-production, the equivalent functions $f_{ij}$ are given by~\cite{Beenakker:1995fp}:
\begin{align}
f_{gg} =& \frac{ \pi m_{\tilde{g}}^2 }{ Q^2 } \left\{ \left[ \frac{9}{4} + \frac{ 9 m_{\tilde{g}}^2 } { Q^2 } - \frac{9 m_{\tilde{g}}^4} { Q^4 } \right] \log \left( \frac{1 + \beta}{1 - \beta} \right) - 3 \beta - \frac{ 51 \beta m_{\tilde{g}}^2 } { 4 Q^2 } \right\}\;,\\ 
f_{q\bar{q}} =& \frac{ \pi m_{\tilde{g}}^2 }{ Q^2 } \left\{ \beta \left[ \frac{20}{27} +\frac{16 m_{\tilde{g}}^2} { 9 Q^2 } - \frac{ 8 m_{-}^2 } { 3 Q^2 } + \frac{ 32 m_{-}^4 } { 27 ( m_{-}^4 + m_{\tilde{q}}^2 Q^2 ) } \right] \right.\\\nonumber
 &+\left. \left[ \frac{ 64 m_{\tilde{q}}^2 } { 27 Q^2 } + \frac{ 8 m_{-}^4 } { 3 Q^4 } - \frac{ 16 m_{\tilde{g}}^2 m_{-}^2 }{ 27 Q^2 ( Q^2 - 2 m_{-}^2 ) }  \right]\log \left( \frac{ 1 - \beta - 2 m^2_{-} / Q^2 } { 1 + \beta - 2 m^2_- / Q^2 } \right) \right\}\;,
\end{align}
where now $\beta=\sqrt{1 - 4 m_{\tilde{g}}^2 / Q^2 }$ and $m_-^2$
represents the mass-squared difference between the gluino and the
t-channel squark, $m_{-}^2 = m ^2 _{\tilde{g}} - m ^2 _{\tilde{q}}$.

\bibliography{isrref}
\bibliographystyle{utphys}

\end{document}